%% file: main_arXiv.tex
\documentclass[superscriptaddress,12pt]{revtex4-2}
\usepackage{titlesec}
\titleformat{\section}{\raggedright\bfseries}{\thesection}{1em}{}
\titleformat{\subsection}{\raggedright\bfseries}{\thesubsection}{1em}{}
\usepackage{graphicx}
\usepackage{dcolumn}
\usepackage{bm}
\usepackage{xcolor}
\usepackage{amsmath}

\begin{document}
\title{Effective Graph Resistance as Cumulative Heat Dissipation}
\author{Xiangrong Wang}
\email{xiangrongwang88@gmail.com}
\author{Xin Yu}
\author{Zongze Wu}
\affiliation{College of Mechatronics and Control Engineering, Shenzhen University, Shenzhen, 518060, China}
\author{Yamir Moreno}
\email{yamir.moreno@gmail.com}
\affiliation{Institute for Biocomputation and Physics of Complex Systems, University of Zaragoza, Zaragoza, 50018, Spain}
\affiliation{Department of Theoretical Physics, University of Zaragoza, Zaragoza, 50009, Spain}
\date{\today}

\begin{abstract}
Effective graph resistance is a fundamental structural metric in network science, widely used to quantify global connectivity, compare network architectures, and assess robustness in flow-based systems. Despite its importance, current formulations rely mainly on spectral or pseudo-inverse Laplacian representations, offering limited physical insight into how structural features shape this quantity or how it can be efficiently optimized. Here we establish an exact and physically transparent relationship between effective graph resistance and the cumulative heat dissipation generated by Laplacian diffusion dynamics. We show that the total heat dissipated during relaxation to equilibrium precisely equals the effective graph resistance. This dynamical viewpoint uncovers a natural multi-scale decomposition of the Laplacian spectrum: early-time dissipation is governed by degree-based local structure, intermediate times isolate eigenvalues below the spectral mean, and long times are dominated by the algebraic connectivity. These multi-scale properties yield continuous and interpretable strategies for modifying network structure and constructing optimized ensembles, enabling improvements that are otherwise NP-hard to achieve via combinatorial methods. Our results unify structural and dynamical perspectives on network connectivity and provide new tools for analyzing, comparing, and optimizing complex networks across domains.
\end{abstract}

\maketitle

\section{Introduction}

Effective graph resistance originates from the classical notion of effective resistance in electrical networks and has become a widely used structural metric for complex systems. Because it accounts for \emph{all} reachable paths between pairs of nodes—beyond the shortest paths—it provides a meaningful measure of the effective distance between any two nodes and reflects the overall graph connectivity. Consequently, effective graph resistance has been employed in multiple contexts, including comparing the structure of different networks~\cite{klein93, xiao2003resistance,Effectivegraphresistanc}, performing network sparsification \cite{spielman2008graph,forrow2018functional,kirkley2025fast}, facilitating network searching \cite{latora2001efficient,gounaris2024braess} and optimizing typologies that are prone to cascading failures~\cite{ghosh2008minimizing,wang2014improving}. It has also proven particularly powerful for assessing the robustness of flow-based systems, where flows naturally propagate along multiple parallel routes, such as electrical power grids, transportation networks, and online social networks~\cite{motter2002cascade,wang2017multi,tyloo2018robustness}. These applications include improving the resilience of power, water, and gas infrastructures, metro systems, and communication networks \cite{kurant2006layered,tejedor2017entropy,tejedor2018multiplex}, the optimization of which without properly incorporating multi-pathways typically leads to Braess paradoxical behaviors \cite{cohen1991paradoxical, braess2005paradox,schafer2022understanding,gounaris2024braess}.

Beyond its role as a purely structural descriptor, effective graph resistance is closely connected to random-walk processes on networks through commute times, Kemeny’s constant~\cite{chandra1989electrical,wang2017kemeny}. This constant measures the expected distance between two nodes selected according to the stationary distribution and is notably independent of the initial node of the walk. Extensive research has sought to explain the origin of this invariance~\cite{kemeny1981generalization,levene2002kemeny,hunter2014role}. More recently, effective graph resistance as a structure and dynamics linker, has been shown powerful in constructing brain functional connectivities from structures \cite{abdelnour2014network}, assessing synchrony fragility \cite{tyloo2018robustness}, predicting pandemic \cite{mercier2022effective}, understanding emergence of collective intelligence in groups \cite{horsevadtransition} and compressing neural networks based on pruning \cite{zenil2019causal,balwani2025constructing}, suggesting a broader potential for understanding coordinated behaviors in complex networked systems.

Despite its broad applicability, existing formulations of effective graph resistance rely on the pseudo-inverse of the graph Laplacian or on sums of inverse Laplacian eigenvalues~\cite{klein93,VanMieghem2011}. While mathematically elegant, these spectral expressions provide limited insight into how structural features at different scales contribute to effective graph resistance or how this quantity responds to controlled modifications of network topology. As a result, it remains challenging to continuously modify network links to optimize the effective graph resistance~\cite{ghosh2008minimizing,spielman2008graph,gounaris2024braess}, meaning that achieving global improvements through local adjustments is computationally difficult. More generally, the relationship between network structure and the diffusive dynamics it shapes remains insufficiently understood.

In this work, we introduce a new dynamical perspective on effective graph resistance by establishing an exact correspondence between this structural quantity and the \emph{cumulative heat dissipation} generated by Laplacian diffusion dynamics. Specifically, we show that the effective graph resistance equals the total amount of heat dissipated from the system during relaxation to steady state—equivalently, the area under the trace of the diffusion operator after subtracting the stationary mode. At the microscopic level, the cumulative heat dissipated by an initially activated node is precisely determined by the average effective resistance between that node and all others. This yields a physically transparent interpretation of effective graph resistance, unifying two traditionally separate viewpoints: static structural metrics and dynamical diffusion processes on networks. Note that, henceforth, we refer to heat diffusion for concreteness. Still, the underlying Laplacian dynamics describes the diffusion of any conserved scalar quantity on a network, such as mass, probability, or information, and all results presented here hold generally for linear diffusion processes.

Crucially, the heat-dissipation formulation reveals a multi-scale decomposition of how different regions of the Laplacian spectrum contribute to effective graph resistance. Short diffusion times emphasize local structural properties such as degree; intermediate times highlight eigenvalues below the spectral mean; and long times are dominated by the algebraic connectivity. These insights naturally enable principled topological modification and network-ensemble algorithms that approximate continuous optimization of effective graph resistance—an otherwise NP-hard problem—through dynamical rather than purely combinatorial principles.

\section{Effective graph resistance of networks}

The effective graph resistance is a widely used graph-level indicator for comparing network structures and for assessing or optimizing the robustness of established topologies. Unlike purely local or shortest-path-based metrics, it incorporates \emph{all} reachable paths and their lengths between any pair of nodes, making it particularly appropriate for flow networks in which transport or interaction naturally follows multiple alternative routes.

To formalize the concept, let $G(N,L)$ be an undirected, connected graph with $N$ nodes and $L$ links. Its adjacency matrix $A$ is an $N\times N$ symmetric matrix with entries $a_{ij}\in\{0,1\}$ indicating the presence or absence of a link between nodes $i$ and $j$. The graph Laplacian is defined as $L = \Delta - A$, where $\Delta = \mathrm{diag}(d_i)$ is the diagonal degree matrix with $d_i = \sum_{j=1}^{N} a_{ij}$. The average degree of the graph is $E[D] = 2L/N$. Because $G$ is undirected, $L$ is symmetric and positive semidefinite, with real non-negative eigenvalues. In particular, one eigenvalue equals zero, with normalized eigenvector $u/\sqrt{N}$, where $u$ is the all-ones vector. The Laplacian eigenvalues are ordered as $0 = \lambda_1 \leq \lambda_2 \leq \cdots \leq \lambda_N$ and $ \phi_k $ denotes the eigenvector corresponding to eigenvalue $ \lambda_k $. The second smallest eigenvalue $\lambda_2$ is coined as the algebraic connectivity \cite{fiedler1973algebraic}, reflecting the global connectedness of the graph.

{\bf Effective resistance between node pairs}. The effective resistance $R_{ij}$ measures the voltage drop induced when a unit current is injected at node $i$ and extracted at node $j$ in the corresponding electrical network representation of $G$ (Fig.~1A). In simple electrical circuits with well-defined series and parallel configurations, the effective resistance can be computed by enumerating all parallel paths:
\begin{equation}
\frac{1}{R_{ij}} = \sum_{P_k} \frac{1}{R_{P_k}},
\label{eq:paths}
\end{equation}
where $R_{P_k}$ is the resistance along path $P_k$. In general complex networks, however, enumerating all paths is computationally infeasible due to the combinatorial explosion of intertwined pathways. Therefore, the effective resistance is calculated through the Moore--Penrose pseudo-inverse of the Laplacian:
\begin{equation}
R_{ij} = L^{\dagger}_{ii} + L^{\dagger}_{jj} - 2 L^{\dagger}_{ij}.
\label{eq:pinv}
\end{equation}

{\bf Graph-level effective resistance}. Aggregating pairwise resistances yields the \emph{effective graph resistance}:
\begin{equation}
R_G = \sum_{i=1}^{N} \sum_{j>i} R_{ij}.
\label{eq:rg_def}
\end{equation}
By combining Eqs.~(\ref{eq:pinv}) and (\ref{eq:rg_def}) with spectral properties of the Laplacian, one obtains the well-known eigenvalue formulation:
\begin{equation}
R_G = N \sum_{i=2}^{N} \frac{1}{\lambda_i}.
\label{eq:rg_spectral}
\end{equation}

This expression provides a compact spectral characterization of $R_G$: networks with more numerous and shorter alternative paths (i.e., those supporting strong diffusive or flow processes) have smaller values of $R_G$. This intuition is illustrated in Fig.~1A, where additional parallel routes reduce the effective resistance. However, despite the utility of Eqs.~(\ref{eq:pinv})–(\ref{eq:rg_spectral}), these formulations offer limited insight into the mechanistic relationship between structural organization and the resulting effective graph resistance. In particular, they do not reveal how structural modifications affect $R_G$, nor do they provide a physical interpretation that connects $R_G$ to dynamical processes on the network. In what follows, we address this gap by deriving an exact connection between effective graph resistance and the cumulative heat dissipation generated by diffusion dynamics on networks.

\section{Effective graph resistance as cumulative heat dissipation of diffusion dynamics}

To connect the effective graph resistance with diffusion processes on networks, we begin by describing the Laplacian diffusion dynamics. Let $x(t)$ denote the vector of heat concentrations on the nodes at time $t$ (Fig.~1B), where $x_i(t)$ is the heat value at node $i$. Given an initial condition $x_0 = x(t=0)$, heat flows along edges proportionally to concentration differences:
\begin{equation}
\frac{d x_i(t)}{dt} = - \sum_{j=1}^N a_{ij}\,(x_i - x_j).
\label{eq:node_diff}
\end{equation}
For the whole system, this becomes:
\begin{equation}
\frac{d x(t)}{dt} = -L x(t),
\label{eq:lap_diff}
\end{equation}
whose solution is governed by the diffusion operator $e^{-tL}$, i.e.,:
\begin{equation}
x(t) = e^{-tL} x_0.
\label{eq:solution}
\end{equation}

Because $L$ is symmetric and positive semidefinite, the system converges to the steady state $x(\infty) = \frac{1}{N}u$, where $u$ is the all-ones vector.

{\bf Nodal cumulative heat dissipation}. To reveal the microscopic connection with effective resistance, we consider the illustrative case where diffusion originates from a single node: $x_0 = e_i$, where $(e_i)_i = 1$ and $(e_i)_{k\neq i}=0$. As diffusion progresses, heat spreads from node $i$ to its neighbors and farther nodes through multiple paths. We now define the \emph{remaining heat} at node $i$ at time $\tau$ as the diagonal entry of the diffusion operator:
\begin{equation}
Z_i(\tau) := (e^{-\tau L})_{ii}.
\label{eq:remaining}
\end{equation}

The \emph{cumulative heat dissipation} from node $i$ up to time $t$ is the total amount of heat that has been transferred from node $i$ to the rest of the network:
\begin{equation}
H_i(t) = \int_0^t Z_i(\tau)\, d\tau.
\label{eq:Hi}
\end{equation}

When the system reaches the steady state at time $t^\ast$, no pairs of nodes exchange any heat. Thus, the cumulative heat dissipation equals the integral of the remaining heat above the steady-state value:
\[
H_i(t^\ast) = \int_{0}^{t^\ast} (x_i(\tau) - x_i(\infty))\, d\tau
           = \int_{0}^{\infty} (Z_i(\tau) - \tfrac{1}{N})\, d\tau,
\]
where the upper limit can be replaced by $\infty$ since $Z_i(\tau)$ decays exponentially after $t^\ast$.

Using the spectral decomposition of $L$ and the identity relating effective resistance to eigenpairs (derived in Methods), one obtains the exact relation:
\begin{equation}
\lim_{t\to t^\ast} H_i(t)
    = \frac{1}{N}\sum_{j=1}^N R_{ij}
      - \frac{R_G}{N^2}
    = E[R_{i\cdot}] - \frac{E[R_{ij}]}{2}.
\label{eq:Hi_final}
\end{equation}
Here, $E[R_{i\cdot}]$ is the average effective resistance from node $i$ to all others, and $E[R_{ij}] = 2R_G/(N(N-1))$ is the network-wide average. Equation~(\ref{eq:Hi_final}) provides a physically transparent interpretation: \emph {the total heat dissipated by node $i$ during diffusion encodes its average effective resistance to the network}. It is important to stress that $H_i(t)$ is not simply $x_i(0)-x_i(t)$: the total heat is conserved at all times ($x(t)^\top u = 1$), and heat can temporarily flow back from neighbors. The cumulative integral captures all transfers from node~$i$, not just net loss, revealing the multi-step spreading structure of diffusion paths.

{\bf Graph-level cumulative heat dissipation}. Summing Eq.~(\ref{eq:Hi}) over all nodes yields the total cumulative heat dissipation:
\begin{equation}
H(t) = \int_0^t \sum_{i=1}^{N} (e^{-\tau L})_{ii}\, d\tau.
\label{eq:Ht_def}
\end{equation}
Denoting the \emph{graph-level remaining heat} by:
\begin{equation}
Z(\tau) := \sum_{i=1}^{N} (e^{-\tau L})_{ii}
         = \sum_{i=1}^{N} e^{-\lambda_i \tau},
\label{eq:Z_graph}
\end{equation}
where the second equality follows from the spectral decomposition of $L$, yielding $H(t) = \int_0^t Z(\tau)\, d\tau$. Taking the limit of $H(t)$ when $t\to t^\ast$ leads to our main theoretical result: \emph{when diffusion reaches steady state, the cumulative heat dissipation equals the effective graph resistance scaled by the size of the network}:
\begin{equation}
\lim_{t\to t^\ast} H(t) = \frac{R_G}{N}.
\label{eq:main_result}
\end{equation}

This identity is exact for all connected graphs and provides a new dynamical interpretation of $R_G$ (see the derivation in the Supplementary Information). Compared to the classical eigenvalue expression in Eq.~(4), which weights eigenvalues by $1/\lambda_i$, the heat-dissipation viewpoint expresses $R_G$ as the integral of $e^{-\lambda_i \tau}$ over time. This reformulation makes the contribution of each eigenvalue \emph{time-resolved}, enabling comparisons at different diffusion scales and clarifying which parts of the Laplacian spectrum dominate the effective graph resistance. Figures~2–3 (see also Fig. S1 in the SI) illustrate that for a wide range of synthetic and real networks, the cumulative heat dissipation indeed converges precisely to $R_G/N$.

\section{Multi-scale properties of cumulative heat dissipation}

We now analyze the \emph{multi-scale} character of the cumulative heat dissipation and show how it enables accurate prediction of the effective graph resistance. The key quantity is the graph-level remaining heat,
\[
Z(t) = \sum_{i=1}^{N} (e^{-tL})_{ii} = \sum_{k=1}^{N} e^{-t\lambda_k},
\]
whose rate of change reveals how different regions of the Laplacian spectrum contribute at different diffusion times.

\noindent{\bf Local, intermediate, and global diffusion scales}. From Eq.~(12), the derivative of the remaining heat at node $i$ is
\begin{equation}
\frac{d}{dt} (e^{-tL})_{ii} = - \sum_{k} \lambda_k e^{-t\lambda_k} (\phi_k)^2_{ii}.
\label{eq:nodal_derivative}
\end{equation}
At very small diffusion times, $t\to 0$, the exponential factors satisfy $e^{-t\lambda_k}\approx 1$ for all $k$, and Eq.~(\ref{eq:nodal_derivative}) reduces to a purely \emph{local} structural quantity:
\begin{equation}
\frac{d}{dt} (e^{-tL})_{ii} 
= 
-\sum_k \lambda_k (\phi_k)^2_{ii}
= -d_i,
\qquad t\to 0.
\label{eq:t_small}
\end{equation}
Thus, the initial rate of heat dissipation reflects only the degree of node $i$. For larger, but still moderate, times $t>0$, the exponential terms begin to suppress contributions from large eigenvalues, and the derivative unfolds into the weighted mixture
\begin{equation}
\frac{d}{dt} (e^{-tL})_{ii} 
= 
-\sum_k (\phi_k)^2_{ii}\, \lambda_k e^{-t\lambda_k}.
\label{eq:t_mid_general}
\end{equation}
This establishes the transition from a local regime (dominated by high-frequency modes) to an intermediate regime where only eigenvalues $\lambda_k$ below a certain threshold contribute significantly.

\noindent{\bf Graph-level multi-scale structure}. At the graph level, differentiating $Z(t)$ yields
\begin{equation}
Z'(t) = \frac{d}{dt} \sum_{i=1}^{N} (e^{-tL})_{ii}
=
-\sum_{k=1}^{N} \lambda_k e^{-t\lambda_k}.
\label{eq:Zprime}
\end{equation}
Additionally, to quantify how different eigenvalues contribute to the slope of $Z(t)$, we examine the sensitivity of $|Z'(t)|$ to each eigenvalue:
	\begin{equation}
		\frac{\partial |Z'(t)|}{\partial \lambda_k}
		= e^{-t\lambda_k}(1 - t\lambda_k),
		\label{eq:partialZ}
	\end{equation}
	which yields the critical threshold
	\[
	\frac{\partial |Z'(t)|}{\partial \lambda_k} \begin{cases}
		\ge 0 & \text{if } t \le 1/\lambda_k, \\
		< 0 & \text{if } t > 1/\lambda_k.
	\end{cases}
	\]
	Thus, at early times, \emph{all} eigenvalues contribute, whereas at intermediate times, only eigenvalues satisfying $\lambda_k < 1/t$ contribute positively. Finally, at late times, only $\lambda_2$ remains active. This result provides a natural spectral filtering mechanism. Three characteristic diffusion regimes emerge naturally:

\begin{itemize}
\item [i)]{\it Local regime} ($t \rightarrow 0$): 
All eigenvalues contribute similarly since $e^{-t\lambda_k}\approx 1$, giving $Z'(0) = -\sum_{k}\lambda_k = -E[D]\,N$. Therefore, the heat dissipation reflects only the total number of edges.

\item [ii)]{\it Intermediate $t$-regime}:  In this regime, high eigenvalues are exponentially suppressed, isolating the contribution of eigenvalues \emph{below the average}. If one defines $t_1 = \frac{1}{E[\lambda]} = \frac{1}{E[D]}$, then for $t>t_1$, only terms with $\lambda_k < E[\lambda]$ remain relevant.

\item [iii)]{\it Global regime ($t\gg 1/\lambda_2$)}: The smallest nonzero eigenvalue $\lambda_2$ dominates because all other modes have decayed, namely, $Z'(t) \approx -\lambda_2\, e^{-t\lambda_2}, \qquad t\gg 1/\lambda_2$.
\end{itemize}

These transitions are illustrated in Fig. \ref{fig:multiscal_diffused_heat}\textbf{A}, where $Z(t)$ evolves from a rapid degree-driven decay to a slow global decay governed by $\lambda_2$. We then analyze the transition from the local scale to the intermediate scale, i.e., properties around
$t_1 = 1/E[\lambda] = 1/E[D]$. For $t < t_1$, all eigenvalues contribute in a similar manner to the
decay of $Z(t)$, so the effective graph resistance $R_G$ is determined by the full spectrum. For
$t > t_1$, however, the contribution of large eigenvalues is exponentially suppressed and the
dynamics is effectively governed by the subset of eigenvalues \(\lambda_k^{\text{below}}\) that lie
\emph{below} the average, i.e., $\lambda_k^{\text{below}} < E[\lambda] = E[D]$. Eigenvalues above the average become negligible for predicting the effective graph resistance. Moreover, the area of $Z(t)$ from time $0$ to $t_1$ is almost indistinguishable across different
graphs with the same $(N,L)$, and this similarity is even more pronounced when the degree
sequence is fixed. For $t > t_1$, the predictive contribution to $R_G$ is captured by
\begin{equation}
R_G^{\text{pre}} \propto \sum_k \left( \frac{1}{\lambda_k^{\text{below}}} - \frac{1}{E[D]} \right),
\label{eq:RG_pre_intermediate}
\end{equation}
where the sum runs over eigenvalues \(\lambda_k^{\text{below}} < E[\lambda]\). As shown in
Fig.~\ref{fig:multiscal_diffused_heat}, the effective graph resistance is well predicted by
Eq.~\eqref{eq:RG_pre_intermediate}.

To further analyze the shift from the intermediate scale to the global scale, i.e., properties
around $t_2 = 1/\lambda_2$, we consider the following predictive form for $R_G$:
\begin{equation}\label{eq:rectangle_area}
R_G^{\text{pre}} \propto
\begin{cases}
\displaystyle
\sum_{\lambda_k < E[\lambda],\, \lambda_k \neq \lambda_2}
\left( \frac{1}{\lambda_k^{\text{below}}} - \frac{1}{E[\lambda]} \right),
& \text{if } \displaystyle\sum_k e^{-\lambda_k^{\text{below}}/\lambda_2} \gg 1, \\[2.0ex]
\displaystyle
\frac{1}{\lambda_2} - \frac{1}{E[\lambda]},
& \text{otherwise.}
\end{cases}
\end{equation}
The rationale is as follows. There is a transition in the contribution of the eigenvalues below
the average when we examine how the predictive quantity depends on $\lambda_2$, that is,
on $\partial R_G^{\text{pre}} / \partial (1/\lambda_2)$. A representative form of this dependence is
\begin{equation}\label{eq:transition_derivative}
\frac{\partial}{\partial 1/\lambda_2}
\sum_{\lambda_k<E[\lambda]} \left( \frac{1 - e^{-\lambda^{\text{below}}_k/\lambda_2}}{\lambda^{\text{below}}_k} - \frac{1}{E[\lambda]} \right)
= 1 - e^{-1}+\sum_{\lambda_k<E[\lambda]}e^{-\lambda^{\text{below}}_k/\lambda_2}.
\end{equation}
When
\begin{equation}\label{eq:synergy_condition}
\sum_{k} e^{-\lambda^{\text{below}}_k/\lambda_2} \gg 1,
\end{equation}
increasing $\lambda_2$ synergistically accelerates the decay of the entire band of eigenvalues
\(\lambda_k^{\text{below}}\), leading to a substantially smaller $R_G$. In contrast, when
condition~\eqref{eq:synergy_condition} is not satisfied, $\lambda_2$ alone dominates the
prediction of $R_G$. Although each \(\lambda_k^{\text{below}}\) determines the curvature of
$Z(t)$ in any given time slice, there is a genuine \emph{acceleration effect} between $\lambda_2$
and the other eigenvalues \(\lambda_k^{\text{below}}\) that governs the transition from the
intermediate to the global diffusion regime.

\section{Cumulative-heat–based continuous network ensemble optimization}

Optimizing the topology of a network to minimize its effective graph resistance is, in general, an NP-hard problem. Greedy methods that iteratively search for the “best” link to add or rewire are computationally expensive, scale poorly with system size, and frequently get trapped in suboptimal configurations. Moreover, the spectral expression of $R_G$ in Eq.~(\ref{eq:rg_spectral}) does not reveal which structural modifications are most effective at different topological scales. The multi-scale structure uncovered in the previous section provides a principled alternative: by examining the decay of the cumulative heat dissipation across time, one can obtain a continuous and interpretable representation of how structural changes affect the entire Laplacian spectrum, and consequently the effective graph resistance.

Consider the remaining heat $Z(t) = \sum_{i=1}^{N} (e^{-tL})_{ii} = \sum_{k=1}^{N} e^{-t\lambda_k}$. At $t=0$, the diffusion process yields $Z(0) = N$, $Z'(0) = -\sum_{k=1}^{N} \lambda_k = -2L$, which are fixed by the number of nodes and links. Hence, all networks with the same $(N,L)$ share an identical local-scale anchor: the initial decay of $Z(t)$ for $t\lesssim 1/E[D]$. Networks with identical degree sequences share an even tighter local-scale anchor, as $\sum_i d_i$ is fixed. Structural optimization therefore cannot proceed through manipulating the early-time behavior of $Z(t)$: all such networks are nearly indistinguishable in this regime.

As shown earlier, diffusion times $t \in [t_1,t_2]$ isolate the eigenvalues below the average $\lambda_k^{\text{below}} < E[\lambda]=E[D]$,
because large eigenvalues are exponentially suppressed. Consequently, modifying the structure such that these eigenvalues move closer to one another (i.e., tightening the low-frequency band) reduces the intermediate-scale contribution to $R_G$. In this regime, the predictive form
\[
R_G^{\mathrm{pre}} \propto 
\sum_{\lambda_k < E[\lambda]}
\left(
\frac{1}{\lambda_k^{\text{below}}} - \frac{1}{E[\lambda]}
\right),
\]
suggests that optimal rewiring strategies should aim to: i) compress the low-frequency spectrum around its mean, ii) eliminate outlier eigenvalues below $E[\lambda]$, and iii), promote mesoscopic structural regularity. In practice, these transformations correspond to reducing structural bottlenecks, increasing the homogeneity of connectivity across meso-scale groups, and mitigating large degree fluctuations that generate low-frequency spread. For example, for networks (of $ N $ nodes and $ L $ links) generated an stochastic block model, we consider two blocks with half size $ N/2 $, inter and intra connected with probability $ p_{\text{intra}} = p/2-\Delta p $, $ p_{\text{inter}} = p/2+\Delta p $. When tuning $ \Delta p $, we continuously optimize the effective graph resistance by reducing bottlenecks, increasing meso-scale homogeneity, and compacting low-frequency modes (see, Fig. S4 in the SI).

Finally, as the diffusion time enters the global regime $t > t_2 = 1/\lambda_2$, the smallest nonzero eigenvalue $\lambda_2$ dominates the decay of $Z(t)$. Increasing $\lambda_2$ therefore reduces the late-time contribution to $R_G$. However, in the previous section, we showed that this contribution is not solely determined by $\lambda_2$. When the synergy condition
\[
\sum_k e^{-\lambda_k^{\text{below}}/\lambda_2} \gg 1
\]
is satisfied, adjusting $\lambda_2$ accelerates the decay of \emph{multiple} low-frequency modes, not just its own. This “acceleration effect’’ produces larger gains in reducing $R_G$ than changing $\lambda_2$ alone. Thus, when the synergy condition is satisfied, the global-scale optimization problem should aim at elevating $\lambda_2$ as much as possible, and structuring the graph so that $\lambda_2$ strongly influences the entire low-frequency band. This corresponds to creating networks with a high algebraic connectivity, tightly clustered low-frequency eigenvalues, and minimal spectral gaps among the modes below $E[\lambda]$.

Figure~\ref{fig:Rg_predict_by_structures} shows that for a fixed number of nodes and links $(N,L)$ (for example, $N=500$ and $E[D]=4$), the possible network structures span a wide range of well-known classes: from circulant regular graphs to Small-World networks, to Erdős–Rényi random graphs, to Scale-Free networks, and finally to community-rich structures. A stricter configuration arises when the degree distribution is also fixed: in that case, the initial behavior of the remaining heat is identical, and rewiring affects only the shift of eigenvalues from above-average to below-average values, making the low-frequency band more compact. This can be visualized through the bulk of the eigenvalue distribution. Across these diverse structures, the effective graph resistance encodes richer information than conventional structural descriptors. Networks with identical average degree, degree variance, clustering coefficient, average shortest-path distance, or eigenratio $\lambda_N/\lambda_2$ may nevertheless exhibit markedly different values of $R_G$. In particular, since the average shortest-path length between uniformly chosen node pairs is always larger than the average effective resistance distance,
\[
E[s] > \frac{2R_G}{N(N-1)} = E[R_{ij}],
\]
the effective graph resistance cannot be adequately explained using standard structural metrics alone.

However, the multi-scale properties of cumulative heat dissipation enable continuous and interpretable strategies for optimizing the effective graph resistance. Minimizing $R_G$ requires increasing the smallest nonzero eigenvalue $\lambda_2$ (the algebraic connectivity), which reflects the global connectedness of the graph. Yet, once $\lambda_2$ is sufficiently large, changes in $\lambda_2$ alone no longer determine $R_G$ (Fig.~6 and Fig.~S3 in the SI). In this regime, further reductions of the effective graph resistance require decreasing the spread between the below-average eigenvalues $\lambda_k^{\text{below}}$ and the mean eigenvalue $E[\lambda]$. This corresponds to specific classes of structural configurations where both high algebraic connectivity and a compact low-frequency spectrum are jointly realized. The same principles extend naturally to real-world networks such as power grids, water or gas distribution systems, transportation networks, and online social networks, as shown in Fig. S1 of the SI. Using the multi-scale behavior of cumulative heat dissipation, one can map these systems in terms of their second- and third-smallest eigenvalues, revealing characteristic spectral regimes. In particular, certain network configurations may permit rapid epidemic-like spreading (due to a small $\lambda_2$) while simultaneously inhibiting diffusion-like transport or flow (due to a large $R_G$ or a short characteristic diffusion time $\tau_c$). These spectral signatures expose structural vulnerabilities and can guide targeted modifications, as illustrated in Fig. S3 of the SI.

\section{Conclusions}

We have introduced a new perspective on the effective graph resistance by establishing an exact and physically transparent connection with the cumulative heat dissipation generated by Laplacian diffusion processes. This dynamical viewpoint generalizes the classical pseudo-inverse and spectral formulations and provides a direct physical interpretation: the effective graph resistance equals the total amount of heat dissipated as the network relaxes from an initially localized excitation to its steady state. In this sense, the effective graph resistance bridges two traditionally distinct domains $-$static structural metrics and dynamical diffusion processes on networks.

Beyond this conceptual contribution, our analysis reveals a natural multi-scale structure in the cumulative heat dissipation, driven by different regions of the Laplacian spectrum. Local diffusion times reflect degree-based structural properties; intermediate times isolate eigenvalues below the spectral mean; and global times are governed by the algebraic connectivity $\lambda_2$. These multi-scale insights directly inform how structural modifications shape the effective graph resistance and, crucially, allow the design of continuous topological optimization strategies that would otherwise be NP-hard using purely combinatorial approaches.

The framework developed here enables the construction of diffusion-guided network ensembles that maintain local structural constraints while progressively optimizing intermediate- and global-scale spectral features. This includes generating “light-model’’ ensembles that preserve early-time diffusion patterns (such as those relevant in echo-chamber formation in social networks \cite{cinelli2021echo}), as well as “intertwined’’ or higher-order ensembles \cite{ferraz2024contagion} that enhance global diffusivity. Moreover, the approach identifies structural regimes (below the synergy condition) where networks are highly susceptible to epidemic-like spreading yet inefficient in facilitating diffusion-like transport, offering practical insights into the vulnerabilities of real systems.

Overall, by connecting effective graph resistance to cumulative heat dissipation, we open new avenues for understanding, comparing, and optimizing the structure and function of complex networks across disciplines. This work provides both theoretical foundations and practical tools for analyzing robustness, flow efficiency, and structural reconfiguration in a wide range of natural and engineered systems.

\section{Methods}\label{sec:Method}

Decomposing the effective resistance via eigen-spectrum decomposition yields
\begin{equation}\label{key}
	R_{ij} = (e_i-e_j)^TL^\dagger(e_i-e_j) =\sum_{k} \frac{1}{\lambda_{k}}\left(\left(\phi_k\right)_i-\left(\phi_k\right)_j\right)^2
\end{equation}
Summing over other nodes $ j $ arrives at
\begin{equation}\label{key}
	\sum_j R_{ij} = N\sum_{k}\frac{1}{\lambda_{k}}\left(\phi_k\right)_i^2 + \sum_{k}\frac{1}{\lambda_{k}}
\end{equation}
from which 
\begin{equation}\label{eq:inverse_eigenvector_sum}
	\sum_{k}\frac{1}{\lambda_{k}}\left(\phi_k\right)_i^2  = \frac{\sum_j R_{ij} }{N} - \frac{R_G}{N}
\end{equation}

The remaining heat at node $ i $ at time $ \tau $ can be decomposed by eigen-spectrum as follows
\begin{equation}\label{key}
\left(	e^{-\tau L}\right)_{ii} = \sum_{k} e^{-\tau \lambda_{k}}\left(\phi_k\right)_i^2
\end{equation}
Integrating over time $ \tau $ till time $ t $ yields
\begin{equation}\label{key}
H_i(t) = 	\int_{0}^{t}\left(	e^{-\tau L}\right)_{ii}d\tau =  \sum_{k} \left(\phi_k\right)_i^2 \int_{0}^{t}e^{-\tau \lambda_{k}}d\tau =  \sum_{k} \frac{1}{\lambda_{k}}\left(\phi_k\right)_i^2 - \sum_{k} \frac{e^{-t\lambda_{k}}}{\lambda_{k}}\left(\phi_k\right)_i^2 
\end{equation}
As the system relaxes to the steady state at time  $t^* $, we arrive at
\begin{equation}\label{key}
	\lim_{t \rightarrow t^*}H_i(t)=\int_{0}^{t^*}\left(e^{-\tau L}\right)_{ii}d\tau = \sum_{k} \frac{1}{\lambda_{k}}\left(\phi_k\right)_i^2
\end{equation}
Combing with Eq. \eqref{eq:inverse_eigenvector_sum}, we establish that
\begin{equation}\label{key}
	\lim_{t \rightarrow t^*}H_i(t)= \frac{\sum_j R_{ij} }{N} - \frac{R_G}{N}
\end{equation}

\newpage

\begin{figure}
	\centering
	\includegraphics[width=\linewidth]{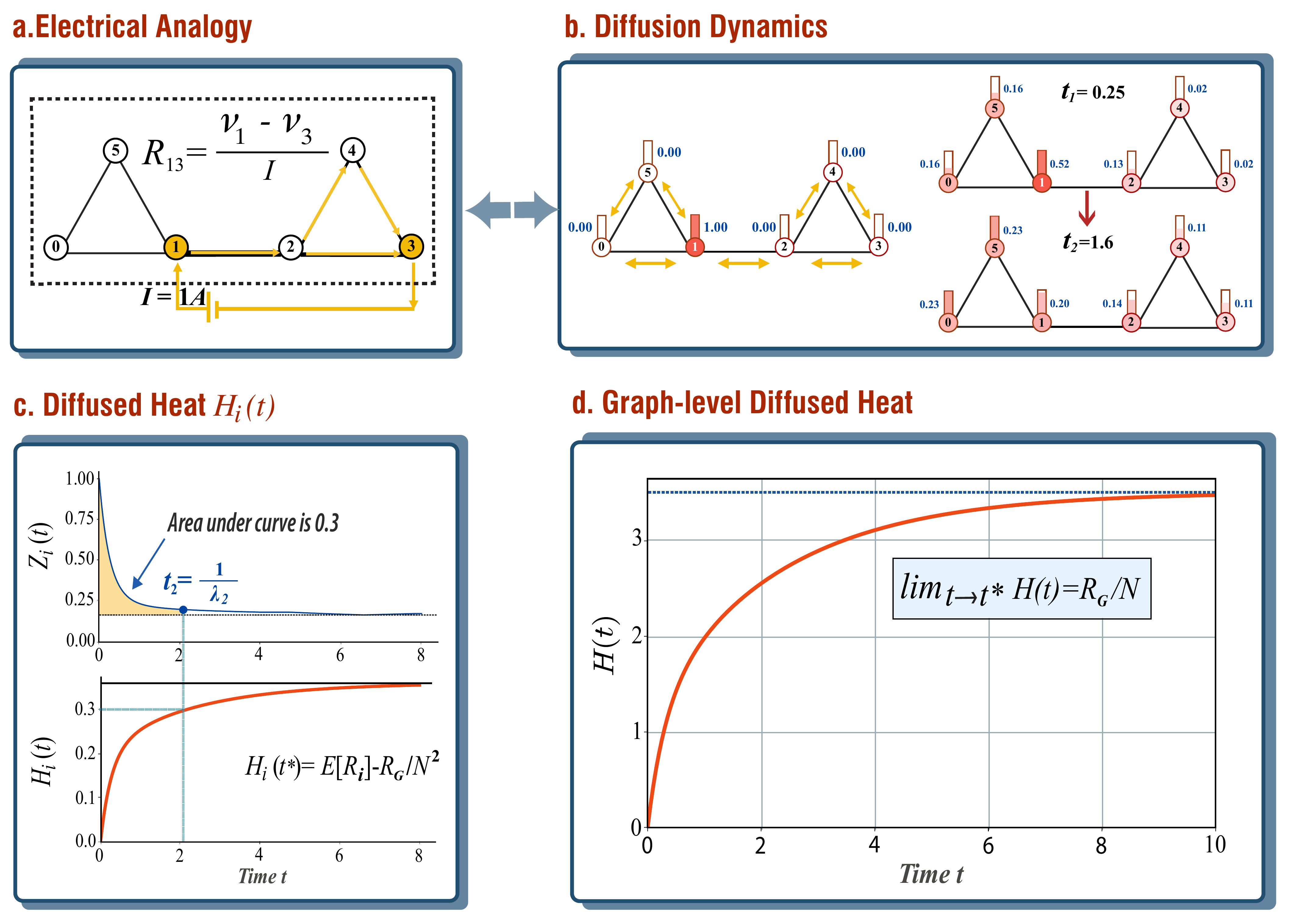}
\caption{\textbf{Effective graph resistance as the cumulative heat dissipation of diffusion dynamics.} 
Panel A illustrates effective resistance in an electrical network, where a larger number of shorter alternative paths between pairs of nodes results in a lower effective graph resistance. 
Panel B shows diffusion dynamics on a network, where the nodal heat concentration $x(t)$ spreads through multiple paths starting from a single initially activated node, $x_0 = e_1$. 
Panel C shows the cumulative heat dissipation $H_i(t)$ associated with the diffusion process in panel B. As the system relaxes to the steady state, $H_i(t)$ converges to the average effective resistance between node $i$ and the rest of the network. 
Panel D shows that the graph-level cumulative heat dissipation converges to the effective graph resistance scaled by the network size.}
	\label{fig:fig1illus}
\end{figure}

\begin{figure}
	\centering
	\includegraphics[width=\linewidth]{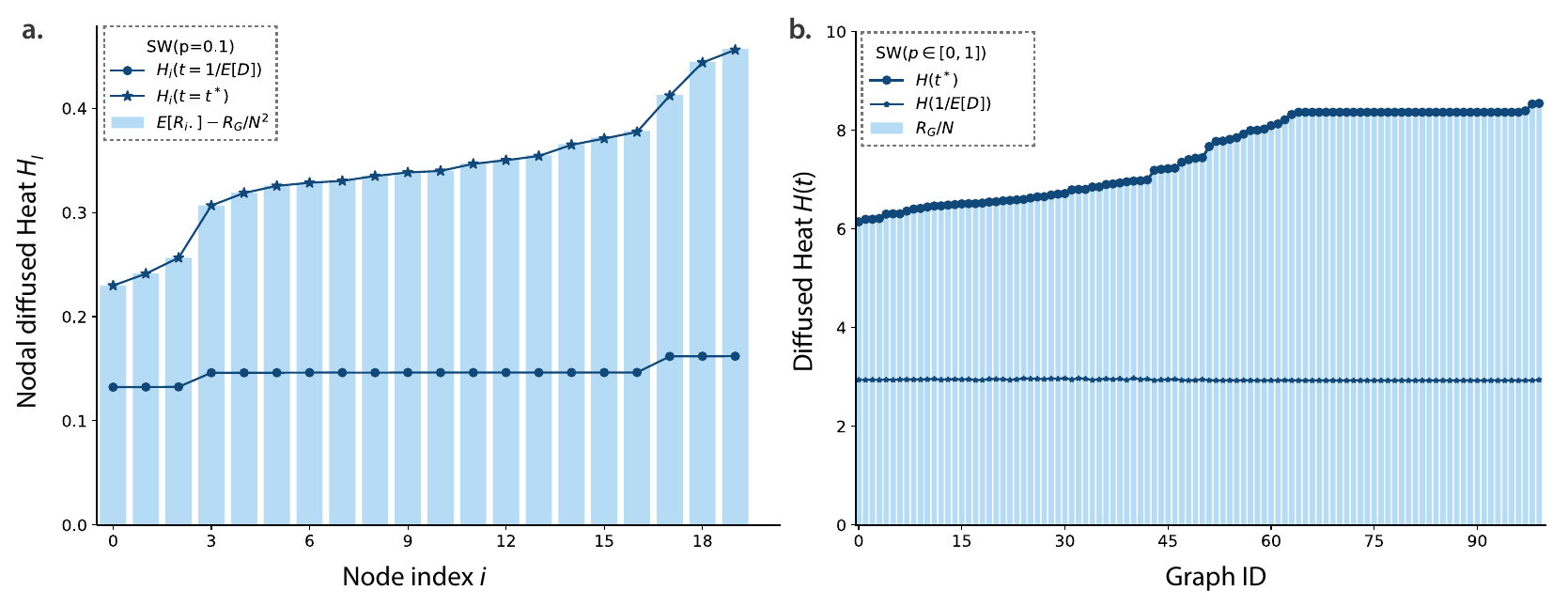}
\caption{\textbf{Cumulative heat dissipation $H(t)$ of diffusion dynamics equals the effective graph resistance $R_G/N$, i.e., $\lim_{t\rightarrow t^*} H(t) = R_G/N$.} 
Panel \textbf{A} shows the nodal average effective resistance (bar height), while the solid line with markers indicates the cumulative heat dissipation $H_i(t^*)$ of node $i$ upon relaxation to the steady state. 
Panel \textbf{B} shows the graph-level cumulative heat dissipation (solid line) together with the effective graph resistance (bar height). 
Panel \textbf{A} corresponds to Watts--Strogatz small-world networks with $N=20$, $E[D]=4$, and link rewiring probability $p=0.1$. 
Panel \textbf{B} corresponds to an ensemble of $100$ small-world networks with $N=20$, $E[D]=4$, and link rewiring probability $p \in [0,1]$.}
	\label{fig:diffused_heat_RG}
\end{figure}

\begin{figure}
	\centering
	\includegraphics[width=0.7\linewidth]{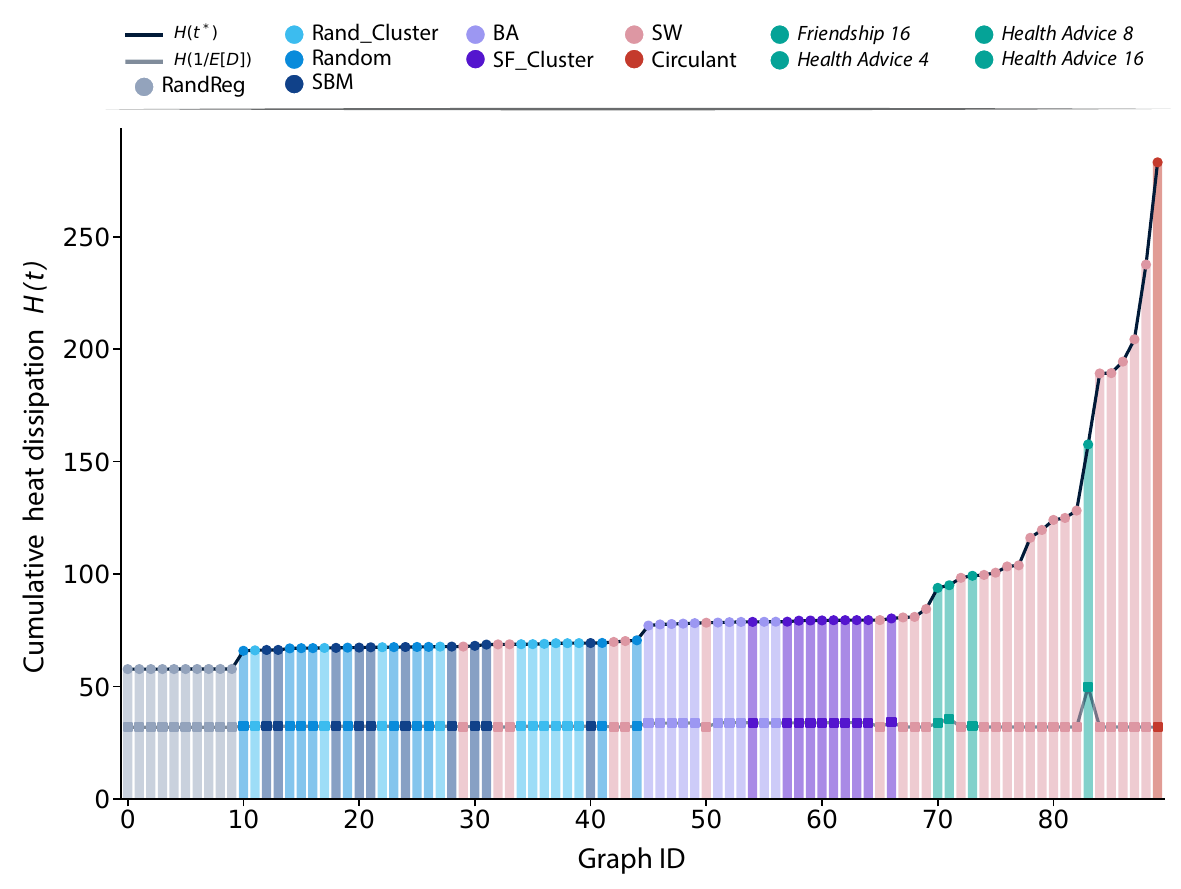}
\caption{Cumulative heat dissipation equals the effective graph resistance $R_G/N$ across a wide range of networks, including circulant regular, small-world, random, scale-free, and community-rich synthetic networks, as well as real-world networks.}
\label{fig:diffusedheatrgrealnet}
\end{figure}

\begin{figure} 
	\centering
	\includegraphics[width=\linewidth]{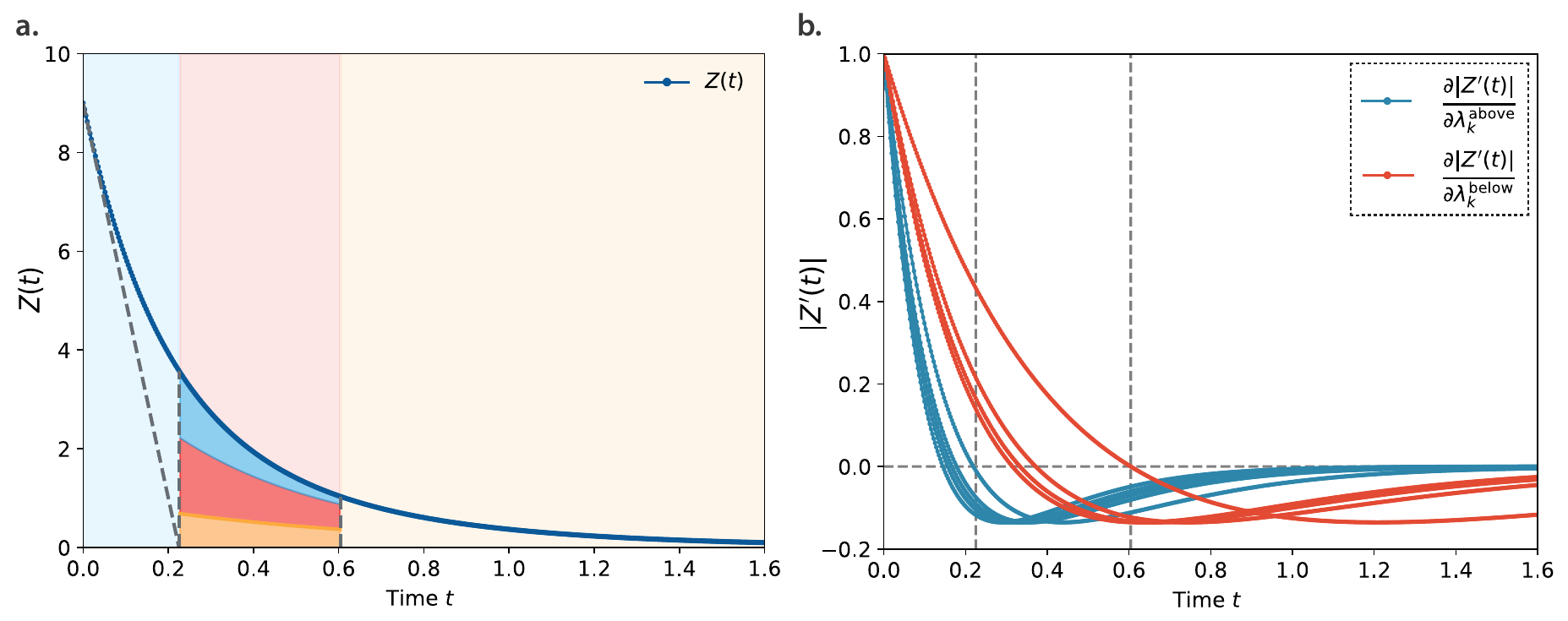}
\caption{\textbf{Multi-scale properties associated with cumulative heat dissipation $H(t)$.} 
Panel \textbf{A} shows the curvature of $Z(t)$ at each time slice, which determines the instantaneous growth rate of the cumulative heat dissipation $H(t)$. 
Panel \textbf{B} shows the contribution of individual Laplacian eigenvalues to the rate of change of $Z'(t)$. 
The shift in eigenvalue contributions naturally reveals distinct diffusion regimes: in the local regime $t \rightarrow 0$, $Z'(0) = -\sum_k \lambda_k$, and $H(t)=\int_0^t Z(\tau)\,d\tau$ reflects only local structural properties such as the number of edges. 
In the intermediate regime $t > 1/E[D]$, only eigenvalues below the spectral mean contribute significantly. 
In the global regime $t > 1/\lambda_2$, the dynamics are dominated by the smallest nonzero eigenvalue $\lambda_2$.}
	\label{fig:multiscal_diffused_heat}
\end{figure}

\begin{figure}
	\centering
	\includegraphics[width=\linewidth]{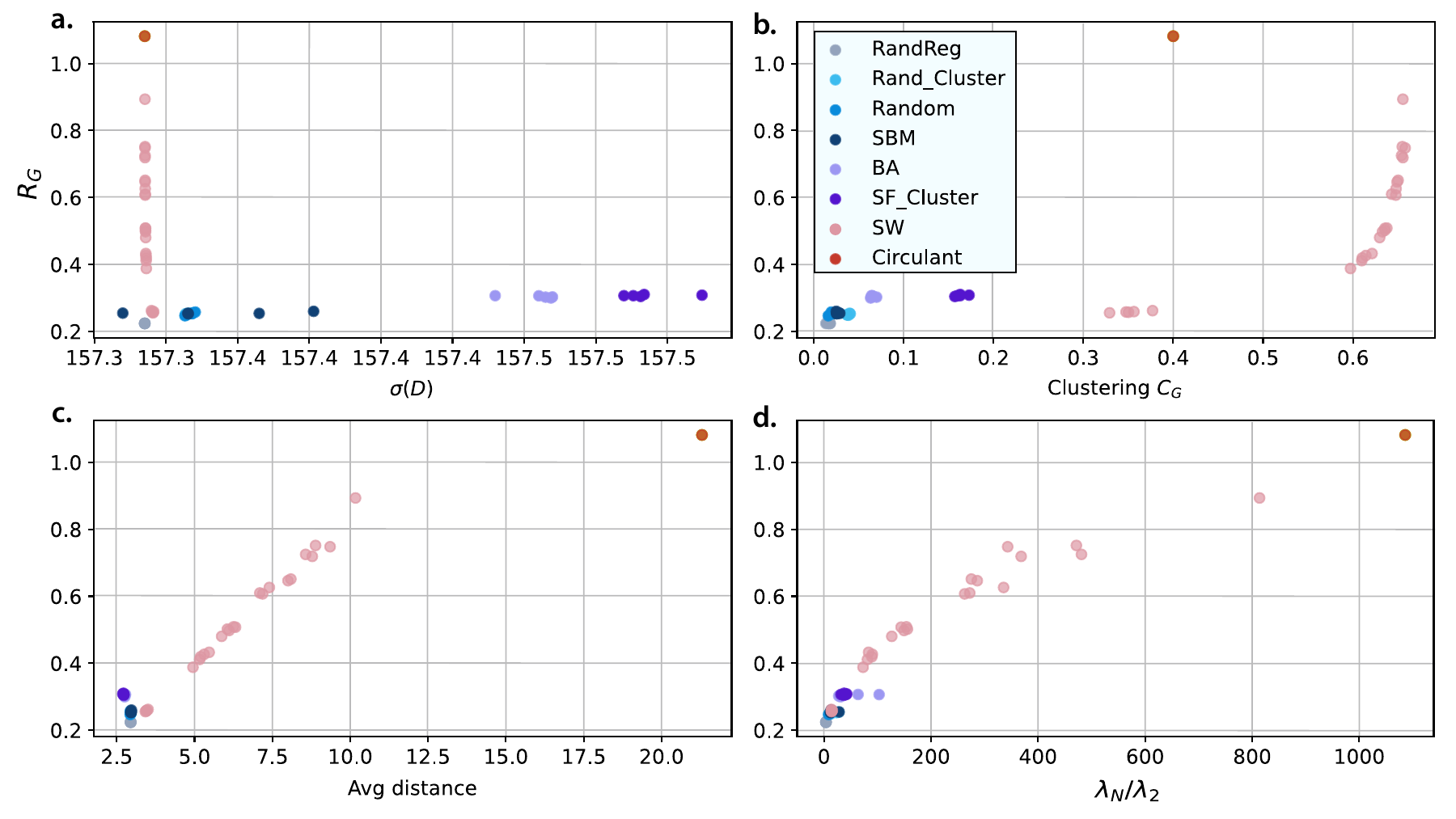}
    \caption{The effective graph resistance encodes richer information than standard structural descriptors, such as average degree, degree standard deviation, clustering coefficient, average shortest-path distance, and the eigenratio $\lambda_N/\lambda_2$.}
\label{fig:Rg_predict_by_structures}
\end{figure}

\begin{figure}
	\centering
	\includegraphics[width=\linewidth]{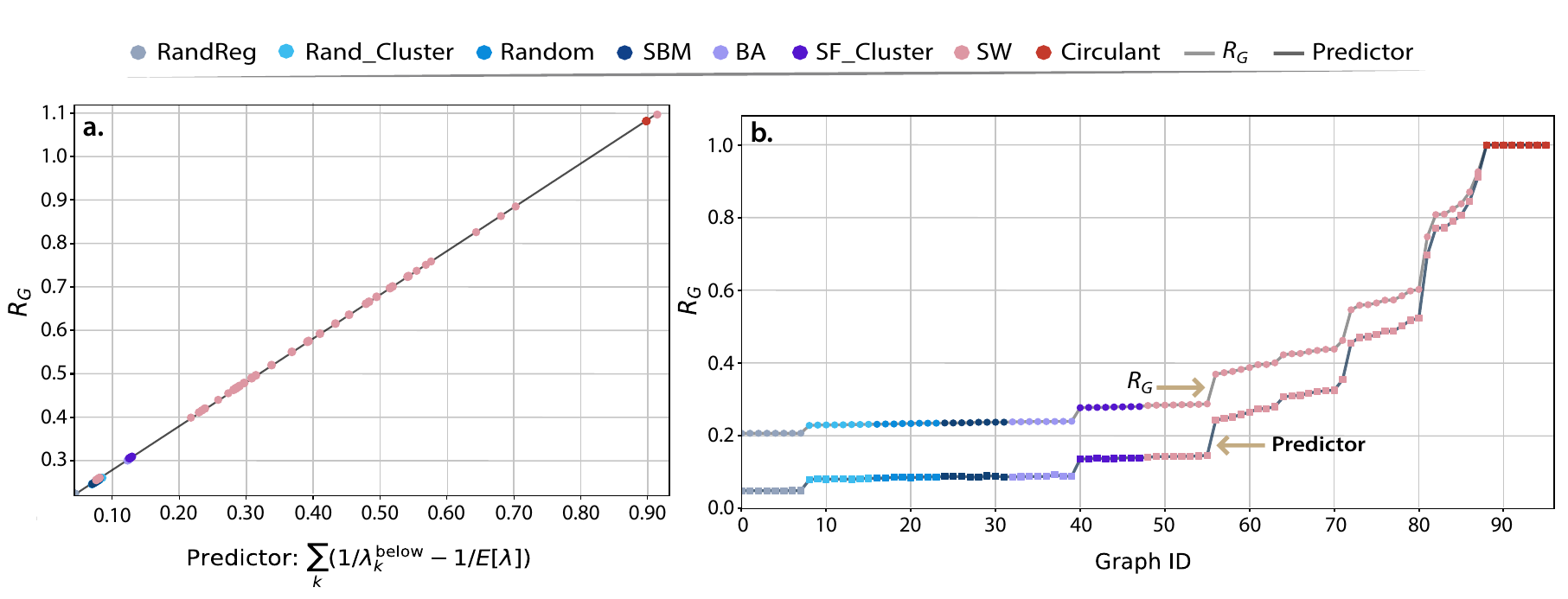}
\caption{\textbf{Multi-scale properties of cumulative heat dissipation accurately predict the effective graph resistance.} 
Panel \textbf{A} shows that the effective graph resistance $R_G$, normalized by the maximum value of $R_G$ for networks with the same $(N,L)$ for visualization purposes, is well predicted by the estimator $R_G^{\text{pre}}$ in Eq.~\eqref{eq:RG_pre_intermediate}. 
Panel \textbf{B} shows that, for graphs sharing the same $(N,L)$, a more compact low-frequency spectrum continuously optimizes the effective graph resistance.}
\label{fig:predictrgn400avgdeg10nocir}
\end{figure}

\clearpage
\appendix
\begin{center}
  \textbf{\large Supplemental Material for "Eﬀective Graph Resistance as Cumulative Heat Dissipation"}
\end{center}
\input{SI_arXiv.tex}

\end{document}

%% file: SI_arXiv.tex
%

\section{Cumulative heat dissipation converges to the effective graph resistance}

Here, we supplement the main text by explicitly connecting the graph-level cumulative heat dissipation to the graph-level effective resistance, i.e., the effective graph resistance. We first formulate the graph-level remaining heat $Z(\tau)$ by summing the nodal remaining heat $Z_i(\tau)$ over all nodes at a given time $\tau$:
\begin{equation}
Z(\tau) := \sum_{i=1}^{N} (e^{-\tau L})_{ii}
= \sum_{i=1}^{N} e^{-\lambda_i \tau}.
\label{S1}
\end{equation}

When the diffusive system reaches the steady state, no pairs of nodes exchange heat. Therefore, the cumulative heat dissipation at the graph level equals the time integral of the remaining heat above the steady state, summed over all nodes:
\begin{equation}
\lim_{t \to t^*} H(t)
= \int_{0}^{t^*} \sum_{i=1}^{N} \left(x_i(\tau) - x_i(\infty)\right) \, d\tau .
\label{S2}
\end{equation}

Given that the diffusive system converges to the steady state $x_i(\infty)=1/N$ and that the remaining heat $Z(\tau)$ decays exponentially beyond the relaxation time $t^*$, the cumulative heat dissipation can be further written as
\begin{equation}
\lim_{t \to t^*} H(t)
= \int_{0}^{t^*} \left(Z(\tau) - 1\right) \, d\tau
= \int_{0}^{\infty} \left( \sum_{i=1}^{N} e^{-\lambda_i \tau} - 1 \right) d\tau .
\label{S3}
\end{equation}

Using the identity $\int_{0}^{\infty} e^{-\lambda_i \tau} d\tau = 1/\lambda_i$, the cumulative heat dissipation is shown to converge exactly to the effective graph resistance scaled by the network size:
\begin{equation}
\lim_{t \to t^*} H(t) = \frac{R_G}{N}.
\label{S4}
\end{equation}

\begin{figure}[!t]
	\centering
	\includegraphics[width=\linewidth]{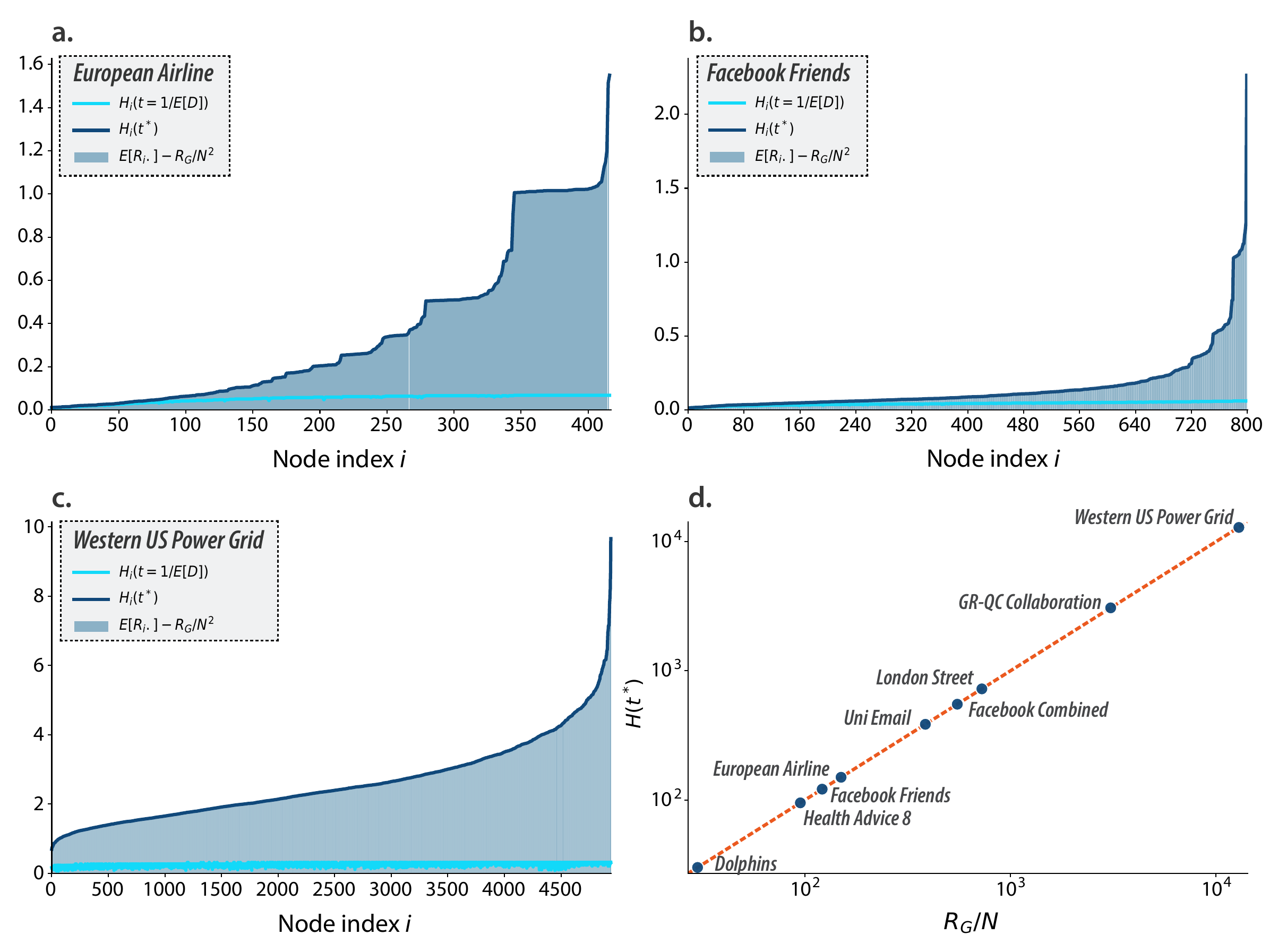}
\caption{Cumulative heat dissipation at both the node level and the graph level converges to the nodal effective resistance and the effective graph resistance, respectively. Panels A--C show that $H_i(t^\ast)$ for each node $i$ converges to the average effective resistance of that node, $E[R_{i\cdot}] - R_G/N^2$. Panel D shows that the graph-level cumulative heat dissipation $H(t^\ast)$ exactly equals $R_G/N$ for real-world flow-oriented networks, including social, transportation, and power networks.}
	\label{fig:figS1}
\end{figure}

Figure \ref{fig:figS1}A-C further illustrate that the cumulative heat dissipation at the microscopic node level converges to the average effective resistance of that node, namely
$H_i(t^*) = E[R_{i\cdot}] - R_G/N^2$, as derived in the main text. Figure \ref{fig:figS1}D shows that the cumulative heat dissipation at the graph level converges exactly to $R_G/N$, in agreement with Eq.~\ref{S4}. The diffusion process is performed on a variety of real-world flow-oriented systems~\cite{lusseau2003bottlenose,leskovec2007graph,chami2017social,guimera2003self,crucitti2006centrality,cardillo2013emergence,leskovec2012learning,sapiezynski2019interaction}, including transportation networks (London street network, European airline network), power grids (Western US power grid), and social networks (Facebook, email, and collaboration networks). Across these diverse network structures and application domains, the cumulative heat dissipation accurately recovers the effective resistance both at the microscopic (nodal) level and at the macroscopic (graph) level.

\section{Multiscale property of cumulative heat dissipation}

We analyze the multiscale properties of the cumulative heat dissipation, which naturally emerge when integrating the graph-level remaining heat $Z(t)$ over time. The key quantity is the curvature change $Z'(t)$ of the remaining heat and how this change is contributed by individual Laplacian eigenvalues. This perspective allows us to formulate the multiscale structure arising from shifts in eigenvalue contributions and to identify characteristic times marking the transition from local to global diffusion regimes.

The contribution of each eigenvalue $\lambda_k$ to the rate of change of $Z(t)$ is quantified by $\partial Z'(t)/\partial \lambda_k$, which reads
\begin{equation}
\frac{\partial Z'(t)}{\partial \lambda_k}
= (\lambda_k t - 1)e^{-\lambda_k t}
\begin{cases}
\le 0, & \text{if } t \le 1/\lambda_k,\\
> 0, & \text{if } t > 1/\lambda_k.
\end{cases}
\label{S5}
\end{equation}

There is a transition in the eigenvalue contribution at the critical time $t = 1/\lambda_k$. A negative contribution implies that increasing this eigenvalue monotonically reduces $Z(t)$, thereby deepening the curvature of $Z(t)$. This negative contribution is further projected onto a lower increase rate of the cumulative heat dissipation $H(t)$ when integrating $Z(t)$ over an infinitesimal time slice. In other words, the eigenvalue $\lambda_k$ remains predictive for the cumulative heat dissipation in the time range $t < 1/\lambda_k$.  

Above the critical time, the contribution becomes positive, implying that the curvature of $Z(t)$ changes only weakly. Accordingly, the eigenvalue has an exponentially diminishing influence on the cumulative heat dissipation $H(t > 1/\lambda_k)$ beyond its characteristic time.

As a result, three characteristic time scales emerge naturally, isolating the roles of different Laplacian eigenvalues: the local regime $t \to 0$, the intermediate time regime $t_1 = 1/E[\lambda] = 1/E[D]$, and the global regime $t_2 = 1/\lambda_2$:
\begin{itemize}
\item \textit{Local regime}: $Z'(0) \simeq \sum_{k=1}^{N} \lambda_k = N E[D]$. All eigenvalues collapse into the total number of edges and jointly determine the cumulative heat dissipation in the short-time limit $H(t \to 0)$.
\item \textit{Intermediate time regime}: The characteristic time $t_1 = 1/E[\lambda]$ filters out high-frequency eigenvalues (those above the spectral mean $E[\lambda]$) in determining the cumulative heat dissipation $H(t)$. In the time window $t > 1/E[\lambda]$ but $t < 1/\lambda_2$, only eigenvalues below the spectral mean remain predictive, as they directly determine the curvature of $Z(t)$ at each time slice.
\item \textit{Global regime}: For times beyond $t_2 = 1/\lambda_2$, the second smallest eigenvalue $\lambda_2$ dominates the cumulative heat dissipation, acting as the slowest decaying mode.
\end{itemize}

\begin{figure}[!t]
	\centering
	\includegraphics[width=\linewidth]{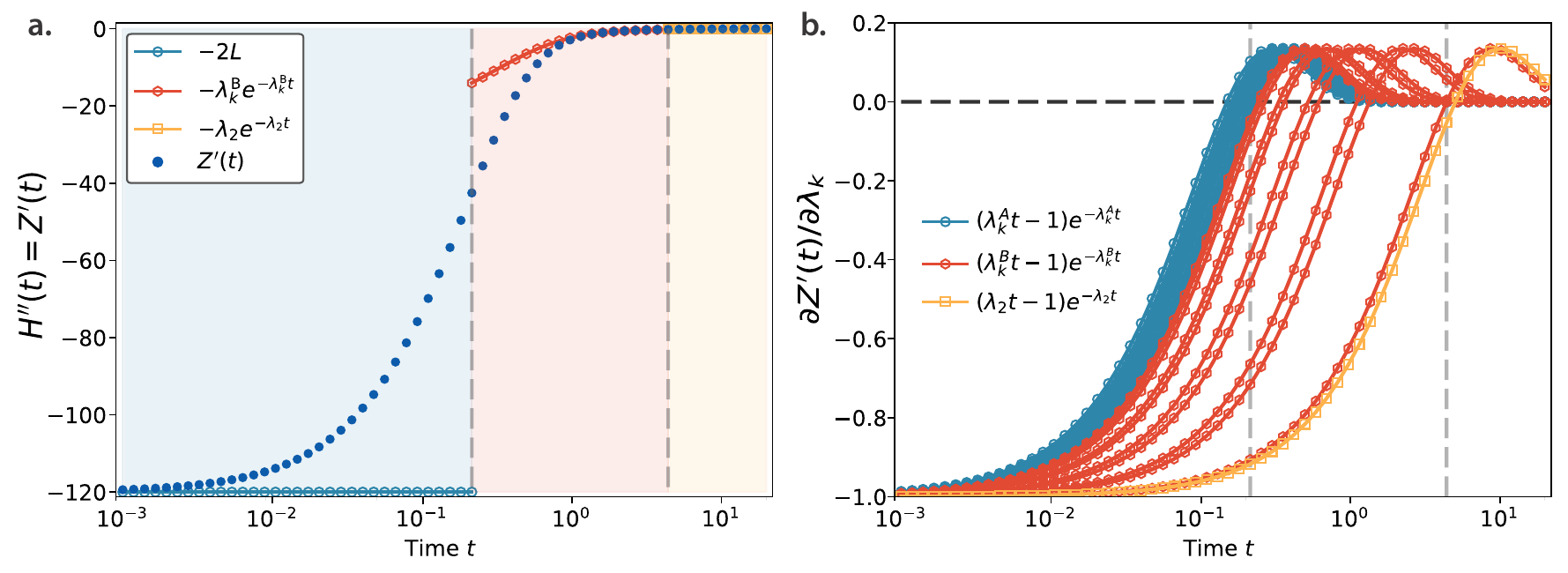}
    \caption{\textbf{Local, intermediate, and global regimes of cumulative heat dissipation.} Panel A shows the rate of change of the remaining heat $Z'(t)$ in the local, intermediate, and global regimes, characterized by the time scales $t_1 = 1/E[D]$ and $t_2 = 1/\lambda_2$ (vertical dotted lines). In each regime, $Z'(t)$ is determined by a filtered composition of Laplacian eigenvalues: all eigenvalues in the local regime, eigenvalues below the spectral mean $\lambda_k^{\text{B}}=\lambda_k^{\text{below}}$ in the intermediate regime, and the second smallest eigenvalue $\lambda_2$ in the global regime. Panel B shows how bundles of eigenvalues transition from negative to positive contributions to $Z'(t)$ across the local, intermediate, and global diffusion regimes.}
	\label{fig:figS2}
\end{figure}

To describe the entire evolution of the cumulative heat dissipation, the multiscale structure can be summarized as
\begin{equation}
H''(t) = Z'(t) =
\begin{cases}
-\sum_k \lambda_k = -2L, & t \to 0,\\
-\sum_k \lambda_k^{\text{below}} e^{-\lambda_k^{\text{below}} t}, & t > 1/E[D],\\
-\lambda_2 e^{-\lambda_2 t}, & t \gg 1/\lambda_2,
\end{cases}
\label{S6}
\end{equation}
where $\lambda_k^{\text{below}}$ denotes eigenvalues below the spectral mean.

The multiscale property of the cumulative heat dissipation reveals the composition of Laplacian eigenvalues governing the diffusion process at different time scales. The local regime is governed by all eigenvalues through the total number of edges, the intermediate regime isolates eigenvalues below the spectral mean, and the global regime is dominated by the second smallest eigenvalue. Correspondingly, the characteristic times mark the transition from locally confined diffusion to mesoscopic spreading and finally to global diffusion.

Figure \ref{fig:figS2} illustrates the multiscale structure of the cumulative heat dissipation formulated in Eq.~\ref{S6}, which arises fundamentally from the pattern shift in eigenvalue contributions.

\section{Multiscale property in predicting the effective graph resistance}

The connection between cumulative heat dissipation and effective graph resistance established above enables the use of multiscale diffusion properties to predict the effective graph resistance. In turn, this provides an interpretable framework for modifying network structure and assembling optimized networks.

To develop a multiscale-driven prediction of the effective graph resistance, we analyze the cumulative heat dissipation at the characteristic times $t_1 = 1/E[D]$ and $t_2 = 1/\lambda_2$, which read
\begin{equation}
H\!\left(t_1=\frac{1}{E[D]}\right)
= \int_{0}^{1/E[D]} \sum_{i=1}^{N} \left(x_i(\tau) - x_i(\infty)\right)\, d\tau
= \int_{0}^{1/E[D]} \sum_{i=2}^{N} e^{-\lambda_i \tau}\, d\tau ,
\label{S7}
\end{equation}
and, analogously, the cumulative heat dissipation at time $t_2$ follows
\begin{equation}
H\!\left(t_2=\frac{1}{\lambda_2}\right)
= \int_{0}^{1/\lambda_2} \sum_{i=1}^{N} \left(x_i(\tau) - x_i(\infty)\right)\, d\tau .
\label{S8}
\end{equation}

The cumulative heat dissipated within the intermediate time window $t \in [1/E[D],\,1/\lambda_2]$ is therefore
\begin{equation}
H(t_2) - H(t_1)
= \int_{1/E[D]}^{1/\lambda_2} \sum_{i=2}^{N} e^{-\lambda_i \tau}\, d\tau
= \sum_{i=2}^{N} \frac{e^{-\lambda_i/E[D]} - e^{-\lambda_i/\lambda_2}}{\lambda_i}.
\label{S9}
\end{equation}

Similarly, the cumulative heat dissipated in the global regime $t \in [1/\lambda_2,\, t^*]$ is given by
\begin{equation}
H(t^*) - H(t_2)
= \sum_{i=2}^{N} \frac{e^{-\lambda_i/\lambda_2} - e^{-\lambda_i t^*}}{\lambda_i}.
\label{S10}
\end{equation}

Further, employing the well-established approximation $1-x \simeq e^{-x}$ for $x<1$ and $e^{-x}\to 0$ for $x\gg 1$, we obtain
\begin{equation}
H(t^*) - H(t_1)
\simeq
\sum_{\lambda_i < E[D]}
\left(
\frac{1}{\lambda_i} - \frac{1}{E[D]}
\right).
\label{S11}
\end{equation}

Since $H(t_1) - H(0) \simeq 1/E[D]$, which is hardly distinguishable for networks with the same number of nodes and links, the predictive factor for the effective graph resistance can be formulated as
\begin{equation}
R_G^{\mathrm{pre}}
\simeq
\sum_{\lambda_i < E[D]}
\left(
\frac{1}{\lambda_i} - \frac{1}{E[D]}
\right).
\label{S12}
\end{equation}
{\bf Synergy condition during the transition from intermediate to global regime}. We further analyze the transition from the intermediate time regime to the global regime, characterized by the second smallest eigenvalue $\lambda_2$. At the characteristic time $t_2 = 1/\lambda_2$, adjusting $\lambda_2$ simultaneously affects the cumulative heat dissipation in both the intermediate and global regimes, but with different weights:
\begin{equation}
\frac{\partial \left(H(t_2) - H(t_1)\right)}{\partial (1/\lambda_2)}
= 1 - e^{-1} + \sum_{\lambda_i < E[D]} e^{-\lambda_i/\lambda_2}
= 1 - e^{-1} + c ,
\label{S13}
\end{equation}
\begin{equation}
\frac{\partial \left(H(t^*) - H(t_2)\right)}{\partial (1/\lambda_2)}
= e^{-1} - \sum_{\lambda_i < E[D]} e^{-\lambda_i/\lambda_2}
= e^{-1} - c ,
\label{S14}
\end{equation}
where we define
\begin{equation}
c := \sum_{\lambda_i < E[D]} e^{-\lambda_i/\lambda_2}.
\label{S15}
\end{equation}

A \emph{synergy condition} is thus identified as $c \gg 1$. Under this condition, $\lambda_2$ plays a significantly compressed role in shaping the cumulative heat dissipation in the global regime [Eq.~\ref{S14}], while the intermediate regime is dominated by the low-frequency spectrum $\lambda_i < E[D]$ other than $\lambda_2$ [Eq.~\ref{S13}]. Consequently, two joint implications follow: (i) the low-frequency spectrum collectively predicts the effective graph resistance, and (ii) compacting $1/\lambda_2$ accelerates the decay of multiple low-frequency modes, yielding larger gains in reducing the effective graph resistance.

The synergy condition also admits a geometric interpretation. Viewing the cumulative heat dissipation as the area of a trapezoid, the height $1/\lambda_2 - 1/E[D]$ and the curve $Z(t)$ act as the two non-parallel sides. Adjusting $\lambda_2$ simultaneously reduces the height and steepens the curvature of $Z(t)$, leading to an accelerated shrinkage of the enclosed area.

\begin{figure}[!t]
	\centering
	\includegraphics[width=1\linewidth]{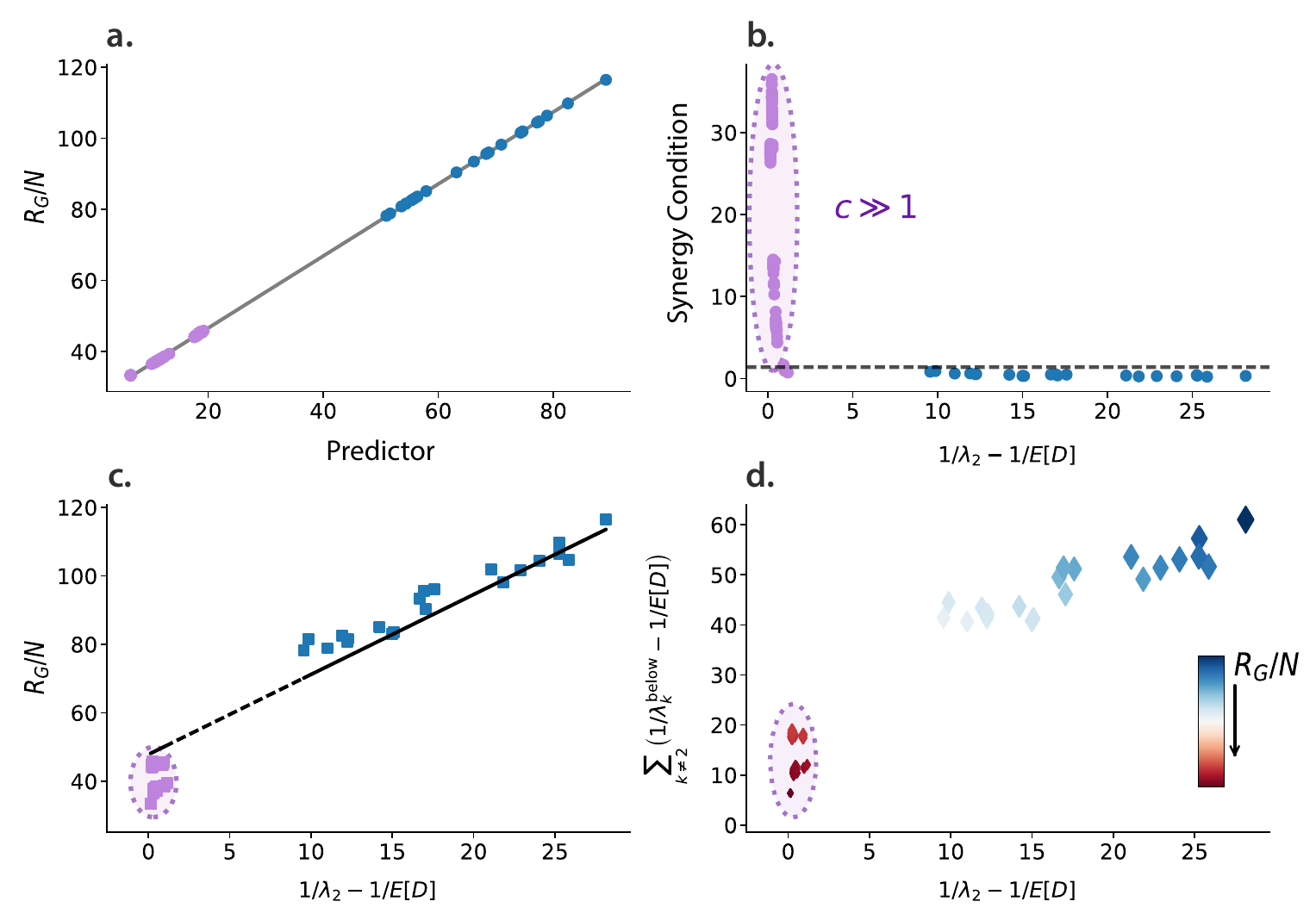}
    \caption{Multiscale-property--driven prediction of the effective graph resistance. Panel A shows that the effective graph resistance is well predicted by the estimator $\sum_{\lambda_i < E[D]} \left(1/\lambda_i - 1/E[D]\right)$. Panels B--D show that, when the synergy condition is satisfied, the effective graph resistance is reduced more strongly than would be predicted by $\lambda_2$ alone. In this synergy regime, other low-frequency modes are also predictive of $R_G$. In contrast, in the regime $c < e^{-1}$, the second smallest eigenvalue $\lambda_2$ alone is predictive of $R_G$.}
	\label{fig:figs3}
\end{figure}

In contrast, another extreme case derived from Eqs.~\ref{S13}–\ref{S14} occurs when
\[
\sum_{\lambda_i < E[D]} e^{-\lambda_i/\lambda_2} < e^{-1}.
\]
In this regime, $\lambda_2$ acts as an outlier and dominantly determines both the intermediate and global heat dissipation. As a result, the second smallest eigenvalue $\lambda_2$ alone predicts the effective graph resistance.

In summary, the multiscale property of cumulative heat dissipation leads to the predictor in Eq.~\ref{S12}, which accurately predicts the effective graph resistance (Fig.~\ref{fig:figs3}A). Moreover, when the synergy condition $c \gg 1$ is satisfied (Fig.~\ref{fig:figs3}B), adjusting $\lambda_2$ induces an acceleration effect that yields additional reductions in the effective graph resistance (Fig.~\ref{fig:figs3}C). When $\lambda_2$ lies in the acceleration regime, low-frequency modes excluding $\lambda_2$ dominate the prediction of the effective graph resistance, whereas when $\lambda_2$ is an outlier, it alone becomes predictive (Fig.~\ref{fig:figs3}D).

\section{Illustration of cumulative-heat--based network optimization}

In the main text, we show that the multiscale property of cumulative heat dissipation enables the formulation of an effective graph resistance predictor, as given in Eq.~\ref{S12}. The predictor highlights that optimization strategies should aim at three aspects: i) compacting the low-frequency spectrum around the mean; ii) eliminating outlier eigenvalues below $E[\lambda]$; and iii) promoting mesoscopic structural regularity. Here, we provide an illustrative example of how such network configurations can be achieved.

We use the stochastic block model to generate networks with $N$ nodes and $L$ links. We consider $k$ blocks of equal size $N_k = N/k$. The probability of having a link between two nodes is $p = L/(N(N-1)/2)$. This construction allows us to design different inter-block and intra-block connection probabilities, defined as $p_{\text{inter}} = p/2 + \Delta p$ and $p_{\text{intra}} = p/2 - \Delta p$, respectively. The tunable parameter $\Delta p \in (-p/2,\, p/2)$ regulates the asymmetry between inter- and intra-block connections, thereby tuning the network structure from a compact low-frequency spectrum to the emergence of outlier small eigenvalues.

\begin{figure}
	\centering
	\includegraphics[width=\linewidth]{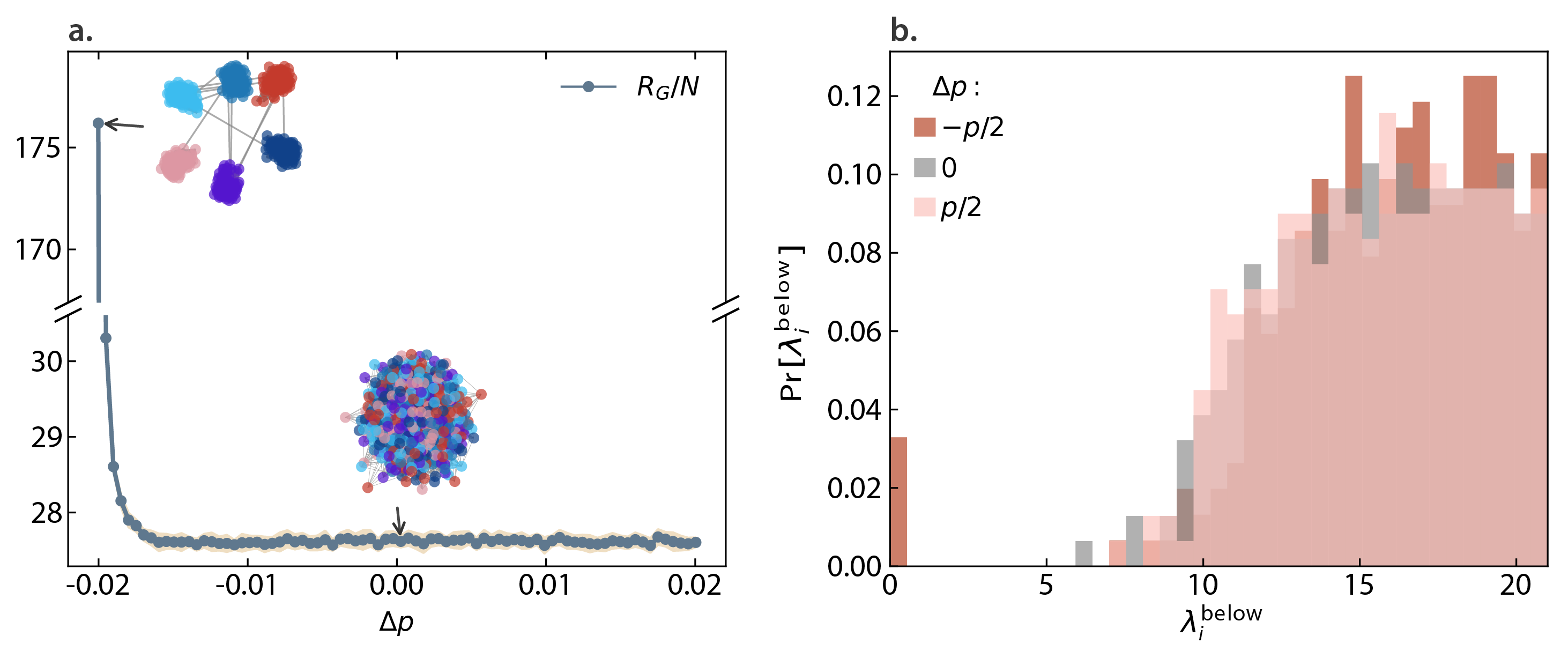}
    \caption{Illustration of cumulative-heat--based optimization by tuning the intra--inter link asymmetry $\Delta p$ in stochastic block models. Panel A shows the scaled effective graph resistance $R_G/N$ as a function of $\Delta p$. Panel B shows the low-frequency spectrum corresponding to $\Delta p = -p/2$ and $\Delta p = 0$. Results are obtained for stochastic block models with $N=500$ nodes and average degree $10$.}
	\label{fig:figs4}
\end{figure}

Figure~\ref{fig:figs4} shows that, for networks consisting of multiple blocks, the effective graph resistance can be continuously optimized by tuning the asymmetry parameter $\Delta p$. This behavior is consistent with the multiscale predictor, which indicates that reducing outliers, promoting mesoscopic regularity, and compacting the low-frequency spectrum lead to lower effective graph resistance.

\newpage

%


%% file: main_arXiv.bbl
\clearpage
\begin{thebibliography}{30}%
\makeatletter
\providecommand \@ifxundefined [1]{%
 \@ifx{#1\undefined}
}%
\providecommand \@ifnum [1]{%
 \ifnum #1\expandafter \@firstoftwo
 \else \expandafter \@secondoftwo
 \fi
}%
\providecommand \@ifx [1]{%
 \ifx #1\expandafter \@firstoftwo
 \else \expandafter \@secondoftwo
 \fi
}%
\providecommand \natexlab [1]{#1}%
\providecommand \enquote  [1]{``#1''}%
\providecommand \bibnamefont  [1]{#1}%
\providecommand \bibfnamefont [1]{#1}%
\providecommand \citenamefont [1]{#1}%
\providecommand \href@noop [0]{\@secondoftwo}%
\providecommand \href [0]{\begingroup \@sanitize@url \@href}%
\providecommand \@href[1]{\@@startlink{#1}\@@href}%
\providecommand \@@href[1]{\endgroup#1\@@endlink}%
\providecommand \@sanitize@url [0]{\catcode `\\12\catcode `\$12\catcode
  `\&12\catcode `\#12\catcode `\^12\catcode `\_12\catcode `\%12\relax}%
\providecommand \@@startlink[1]{}%
\providecommand \@@endlink[0]{}%
\providecommand \url  [0]{\begingroup\@sanitize@url \@url }%
\providecommand \@url [1]{\endgroup\@href {#1}{\urlprefix }}%
\providecommand \urlprefix  [0]{URL }%
\providecommand \Eprint [0]{\href }%
\providecommand \doibase [0]{https://doi.org/}%
\providecommand \selectlanguage [0]{\@gobble}%
\providecommand \bibinfo  [0]{\@secondoftwo}%
\providecommand \bibfield  [0]{\@secondoftwo}%
\providecommand \translation [1]{[#1]}%
\providecommand \BibitemOpen [0]{}%
\providecommand \bibitemStop [0]{}%
\providecommand \bibitemNoStop [0]{.\EOS\space}%
\providecommand \EOS [0]{\spacefactor3000\relax}%
\providecommand \BibitemShut  [1]{\csname bibitem#1\endcsname}%
\let\auto@bib@innerbib\@empty
\bibitem [{\citenamefont {Klein}\ and\ \citenamefont
  {Randi{\'{c}}}(1993)}]{klein93}%
  \BibitemOpen
  \bibfield  {author} {\bibinfo {author} {\bibfnamefont {D.~J.}\ \bibnamefont
  {Klein}}\ and\ \bibinfo {author} {\bibfnamefont {M.}~\bibnamefont
  {Randi{\'{c}}}},\ }\bibfield  {title} {\bibinfo {title} {Resistance
  distance},\ }\href {https://doi.org/10.1007/BF01164627} {\bibfield  {journal}
  {\bibinfo  {journal} {Journal of Mathematical Chemistry}\ }\textbf {\bibinfo
  {volume} {12}},\ \bibinfo {pages} {81} (\bibinfo {year} {1993})}\BibitemShut
  {NoStop}%
\bibitem [{\citenamefont {Xiao}\ and\ \citenamefont
  {Gutman}(2003)}]{xiao2003resistance}%
  \BibitemOpen
  \bibfield  {author} {\bibinfo {author} {\bibfnamefont {W.}~\bibnamefont
  {Xiao}}\ and\ \bibinfo {author} {\bibfnamefont {I.}~\bibnamefont {Gutman}},\
  }\bibfield  {title} {\bibinfo {title} {Resistance distance and laplacian
  spectrum},\ }\href@noop {} {\bibfield  {journal} {\bibinfo  {journal}
  {Theoretical chemistry accounts}\ }\textbf {\bibinfo {volume} {110}},\
  \bibinfo {pages} {284} (\bibinfo {year} {2003})}\BibitemShut {NoStop}%
\bibitem [{\citenamefont {Ellens}\ \emph {et~al.}(2011)\citenamefont {Ellens},
  \citenamefont {Spieksma}, \citenamefont {Van~Mieghem}, \citenamefont
  {Jamakovic},\ and\ \citenamefont {Kooij}}]{Effectivegraphresistanc}%
  \BibitemOpen
  \bibfield  {author} {\bibinfo {author} {\bibfnamefont {W.}~\bibnamefont
  {Ellens}}, \bibinfo {author} {\bibfnamefont {F.~M.}\ \bibnamefont
  {Spieksma}}, \bibinfo {author} {\bibfnamefont {P.}~\bibnamefont
  {Van~Mieghem}}, \bibinfo {author} {\bibfnamefont {A.}~\bibnamefont
  {Jamakovic}},\ and\ \bibinfo {author} {\bibfnamefont {R.~E.}\ \bibnamefont
  {Kooij}},\ }\bibfield  {title} {\bibinfo {title} {Effective graph
  resistance},\ }\href@noop {} {\bibfield  {journal} {\bibinfo  {journal}
  {Linear algebra and its applications}\ }\textbf {\bibinfo {volume} {435}},\
  \bibinfo {pages} {2491} (\bibinfo {year} {2011})}\BibitemShut {NoStop}%
\bibitem [{\citenamefont {Spielman}\ and\ \citenamefont
  {Srivastava}(2008)}]{spielman2008graph}%
  \BibitemOpen
  \bibfield  {author} {\bibinfo {author} {\bibfnamefont {D.~A.}\ \bibnamefont
  {Spielman}}\ and\ \bibinfo {author} {\bibfnamefont {N.}~\bibnamefont
  {Srivastava}},\ }\bibfield  {title} {\bibinfo {title} {Graph sparsification
  by effective resistances},\ }in\ \href@noop {} {\emph {\bibinfo {booktitle}
  {Proceedings of the 14th annual ACM symposium on Theory of computing}}}\
  (\bibinfo {year} {2008})\ pp.\ \bibinfo {pages} {563--568}\BibitemShut
  {NoStop}%
\bibitem [{\citenamefont {Forrow}\ \emph {et~al.}(2018)\citenamefont {Forrow},
  \citenamefont {Woodhouse},\ and\ \citenamefont
  {Dunkel}}]{forrow2018functional}%
  \BibitemOpen
  \bibfield  {author} {\bibinfo {author} {\bibfnamefont {A.}~\bibnamefont
  {Forrow}}, \bibinfo {author} {\bibfnamefont {F.~G.}\ \bibnamefont
  {Woodhouse}},\ and\ \bibinfo {author} {\bibfnamefont {J.}~\bibnamefont
  {Dunkel}},\ }\bibfield  {title} {\bibinfo {title} {Functional control of
  network dynamics using designed laplacian spectra},\ }\href@noop {}
  {\bibfield  {journal} {\bibinfo  {journal} {Physical Review X}\ }\textbf
  {\bibinfo {volume} {8}},\ \bibinfo {pages} {041043} (\bibinfo {year}
  {2018})}\BibitemShut {NoStop}%
\bibitem [{\citenamefont {Kirkley}(2025)}]{kirkley2025fast}%
  \BibitemOpen
  \bibfield  {author} {\bibinfo {author} {\bibfnamefont {A.}~\bibnamefont
  {Kirkley}},\ }\bibfield  {title} {\bibinfo {title} {Fast nonparametric
  inference of network backbones for weighted graph sparsification},\
  }\href@noop {} {\bibfield  {journal} {\bibinfo  {journal} {Physical Review
  X}\ }\textbf {\bibinfo {volume} {15}},\ \bibinfo {pages} {031013} (\bibinfo
  {year} {2025})}\BibitemShut {NoStop}%
\bibitem [{\citenamefont {Latora}\ and\ \citenamefont
  {Marchiori}(2001)}]{latora2001efficient}%
  \BibitemOpen
  \bibfield  {author} {\bibinfo {author} {\bibfnamefont {V.}~\bibnamefont
  {Latora}}\ and\ \bibinfo {author} {\bibfnamefont {M.}~\bibnamefont
  {Marchiori}},\ }\bibfield  {title} {\bibinfo {title} {Efficient behavior of
  small-world networks},\ }\href@noop {} {\bibfield  {journal} {\bibinfo
  {journal} {Physical review letters}\ }\textbf {\bibinfo {volume} {87}},\
  \bibinfo {pages} {198701} (\bibinfo {year} {2001})}\BibitemShut {NoStop}%
\bibitem [{\citenamefont {Gounaris}\ and\ \citenamefont
  {Katifori}(2024)}]{gounaris2024braess}%
  \BibitemOpen
  \bibfield  {author} {\bibinfo {author} {\bibfnamefont {G.}~\bibnamefont
  {Gounaris}}\ and\ \bibinfo {author} {\bibfnamefont {E.}~\bibnamefont
  {Katifori}},\ }\bibfield  {title} {\bibinfo {title} {Braess’s paradox
  analog in physical networks of optimal exploration},\ }\href@noop {}
  {\bibfield  {journal} {\bibinfo  {journal} {Physical Review Letters}\
  }\textbf {\bibinfo {volume} {133}},\ \bibinfo {pages} {067401} (\bibinfo
  {year} {2024})}\BibitemShut {NoStop}%
\bibitem [{\citenamefont {Ghosh}\ \emph {et~al.}(2008)\citenamefont {Ghosh},
  \citenamefont {Boyd},\ and\ \citenamefont {Saberi}}]{ghosh2008minimizing}%
  \BibitemOpen
  \bibfield  {author} {\bibinfo {author} {\bibfnamefont {A.}~\bibnamefont
  {Ghosh}}, \bibinfo {author} {\bibfnamefont {S.}~\bibnamefont {Boyd}},\ and\
  \bibinfo {author} {\bibfnamefont {A.}~\bibnamefont {Saberi}},\ }\bibfield
  {title} {\bibinfo {title} {Minimizing effective resistance of a graph},\
  }\href@noop {} {\bibfield  {journal} {\bibinfo  {journal} {SIAM Review}\
  }\textbf {\bibinfo {volume} {50}},\ \bibinfo {pages} {37} (\bibinfo {year}
  {2008})}\BibitemShut {NoStop}%
\bibitem [{\citenamefont {Wang}\ \emph {et~al.}(2014)\citenamefont {Wang},
  \citenamefont {Pournaras}, \citenamefont {Kooij},\ and\ \citenamefont
  {Van~Mieghem}}]{wang2014improving}%
  \BibitemOpen
  \bibfield  {author} {\bibinfo {author} {\bibfnamefont {X.}~\bibnamefont
  {Wang}}, \bibinfo {author} {\bibfnamefont {E.}~\bibnamefont {Pournaras}},
  \bibinfo {author} {\bibfnamefont {R.~E.}\ \bibnamefont {Kooij}},\ and\
  \bibinfo {author} {\bibfnamefont {P.}~\bibnamefont {Van~Mieghem}},\
  }\bibfield  {title} {\bibinfo {title} {Improving robustness of complex
  networks via the effective graph resistance},\ }\href@noop {} {\bibfield
  {journal} {\bibinfo  {journal} {The European Physical Journal B}\ }\textbf
  {\bibinfo {volume} {87}},\ \bibinfo {pages} {1} (\bibinfo {year}
  {2014})}\BibitemShut {NoStop}%
\bibitem [{\citenamefont {Motter}\ and\ \citenamefont
  {Lai}(2002)}]{motter2002cascade}%
  \BibitemOpen
  \bibfield  {author} {\bibinfo {author} {\bibfnamefont {A.~E.}\ \bibnamefont
  {Motter}}\ and\ \bibinfo {author} {\bibfnamefont {Y.-C.}\ \bibnamefont
  {Lai}},\ }\bibfield  {title} {\bibinfo {title} {Cascade-based attacks on
  complex networks},\ }\href@noop {} {\bibfield  {journal} {\bibinfo  {journal}
  {Physical Review E}\ }\textbf {\bibinfo {volume} {66}},\ \bibinfo {pages}
  {065102} (\bibinfo {year} {2002})}\BibitemShut {NoStop}%
\bibitem [{\citenamefont {Wang}\ \emph
  {et~al.}(2017{\natexlab{a}})\citenamefont {Wang}, \citenamefont {Ko{\c{c}}},
  \citenamefont {Derrible}, \citenamefont {Ahmad}, \citenamefont {Pino},\ and\
  \citenamefont {Kooij}}]{wang2017multi}%
  \BibitemOpen
  \bibfield  {author} {\bibinfo {author} {\bibfnamefont {X.}~\bibnamefont
  {Wang}}, \bibinfo {author} {\bibfnamefont {Y.}~\bibnamefont {Ko{\c{c}}}},
  \bibinfo {author} {\bibfnamefont {S.}~\bibnamefont {Derrible}}, \bibinfo
  {author} {\bibfnamefont {S.~N.}\ \bibnamefont {Ahmad}}, \bibinfo {author}
  {\bibfnamefont {W.~J.}\ \bibnamefont {Pino}},\ and\ \bibinfo {author}
  {\bibfnamefont {R.~E.}\ \bibnamefont {Kooij}},\ }\bibfield  {title} {\bibinfo
  {title} {Multi-criteria robustness analysis of metro networks},\ }\href@noop
  {} {\bibfield  {journal} {\bibinfo  {journal} {Physica A: Statistical
  Mechanics and its Applications}\ }\textbf {\bibinfo {volume} {474}},\
  \bibinfo {pages} {19} (\bibinfo {year} {2017}{\natexlab{a}})}\BibitemShut
  {NoStop}%
\bibitem [{\citenamefont {Tyloo}\ \emph {et~al.}(2018)\citenamefont {Tyloo},
  \citenamefont {Coletta},\ and\ \citenamefont
  {Jacquod}}]{tyloo2018robustness}%
  \BibitemOpen
  \bibfield  {author} {\bibinfo {author} {\bibfnamefont {M.}~\bibnamefont
  {Tyloo}}, \bibinfo {author} {\bibfnamefont {T.}~\bibnamefont {Coletta}},\
  and\ \bibinfo {author} {\bibfnamefont {P.}~\bibnamefont {Jacquod}},\
  }\bibfield  {title} {\bibinfo {title} {Robustness of synchrony in complex
  networks and generalized kirchhoff indices},\ }\href@noop {} {\bibfield
  {journal} {\bibinfo  {journal} {Physical review letters}\ }\textbf {\bibinfo
  {volume} {120}},\ \bibinfo {pages} {084101} (\bibinfo {year}
  {2018})}\BibitemShut {NoStop}%
\bibitem [{\citenamefont {Kurant}\ and\ \citenamefont
  {Thiran}(2006)}]{kurant2006layered}%
  \BibitemOpen
  \bibfield  {author} {\bibinfo {author} {\bibfnamefont {M.}~\bibnamefont
  {Kurant}}\ and\ \bibinfo {author} {\bibfnamefont {P.}~\bibnamefont
  {Thiran}},\ }\bibfield  {title} {\bibinfo {title} {Layered complex
  networks},\ }\href@noop {} {\bibfield  {journal} {\bibinfo  {journal}
  {Physical review letters}\ }\textbf {\bibinfo {volume} {96}},\ \bibinfo
  {pages} {138701} (\bibinfo {year} {2006})}\BibitemShut {NoStop}%
\bibitem [{\citenamefont {Tejedor}\ \emph {et~al.}(2017)\citenamefont
  {Tejedor}, \citenamefont {Longjas}, \citenamefont {Edmonds}, \citenamefont
  {Zaliapin}, \citenamefont {Georgiou}, \citenamefont {Rinaldo},\ and\
  \citenamefont {Foufoula-Georgiou}}]{tejedor2017entropy}%
  \BibitemOpen
  \bibfield  {author} {\bibinfo {author} {\bibfnamefont {A.}~\bibnamefont
  {Tejedor}}, \bibinfo {author} {\bibfnamefont {A.}~\bibnamefont {Longjas}},
  \bibinfo {author} {\bibfnamefont {D.~A.}\ \bibnamefont {Edmonds}}, \bibinfo
  {author} {\bibfnamefont {I.}~\bibnamefont {Zaliapin}}, \bibinfo {author}
  {\bibfnamefont {T.~T.}\ \bibnamefont {Georgiou}}, \bibinfo {author}
  {\bibfnamefont {A.}~\bibnamefont {Rinaldo}},\ and\ \bibinfo {author}
  {\bibfnamefont {E.}~\bibnamefont {Foufoula-Georgiou}},\ }\bibfield  {title}
  {\bibinfo {title} {Entropy and optimality in river deltas},\ }\href@noop {}
  {\bibfield  {journal} {\bibinfo  {journal} {Proceedings of the National
  Academy of Sciences}\ }\textbf {\bibinfo {volume} {114}},\ \bibinfo {pages}
  {11651} (\bibinfo {year} {2017})}\BibitemShut {NoStop}%
\bibitem [{\citenamefont {Tejedor}\ \emph {et~al.}(2018)\citenamefont
  {Tejedor}, \citenamefont {Longjas}, \citenamefont {Passalacqua},
  \citenamefont {Moreno},\ and\ \citenamefont
  {Foufoula-Georgiou}}]{tejedor2018multiplex}%
  \BibitemOpen
  \bibfield  {author} {\bibinfo {author} {\bibfnamefont {A.}~\bibnamefont
  {Tejedor}}, \bibinfo {author} {\bibfnamefont {A.}~\bibnamefont {Longjas}},
  \bibinfo {author} {\bibfnamefont {P.}~\bibnamefont {Passalacqua}}, \bibinfo
  {author} {\bibfnamefont {Y.}~\bibnamefont {Moreno}},\ and\ \bibinfo {author}
  {\bibfnamefont {E.}~\bibnamefont {Foufoula-Georgiou}},\ }\bibfield  {title}
  {\bibinfo {title} {Multiplex networks: A framework for studying multiprocess
  multiscale connectivity via coupled-network theory with an application to
  river deltas},\ }\href@noop {} {\bibfield  {journal} {\bibinfo  {journal}
  {Geophysical Research Letters}\ }\textbf {\bibinfo {volume} {45}},\ \bibinfo
  {pages} {9681} (\bibinfo {year} {2018})}\BibitemShut {NoStop}%
\bibitem [{\citenamefont {Cohen}\ and\ \citenamefont
  {Horowitz}(1991)}]{cohen1991paradoxical}%
  \BibitemOpen
  \bibfield  {author} {\bibinfo {author} {\bibfnamefont {J.~E.}\ \bibnamefont
  {Cohen}}\ and\ \bibinfo {author} {\bibfnamefont {P.}~\bibnamefont
  {Horowitz}},\ }\bibfield  {title} {\bibinfo {title} {Paradoxical behaviour of
  mechanical and electrical networks},\ }\href@noop {} {\bibfield  {journal}
  {\bibinfo  {journal} {Nature}\ }\textbf {\bibinfo {volume} {352}},\ \bibinfo
  {pages} {699} (\bibinfo {year} {1991})}\BibitemShut {NoStop}%
\bibitem [{\citenamefont {Braess}\ \emph {et~al.}(2005)\citenamefont {Braess},
  \citenamefont {Nagurney},\ and\ \citenamefont
  {Wakolbinger}}]{braess2005paradox}%
  \BibitemOpen
  \bibfield  {author} {\bibinfo {author} {\bibfnamefont {D.}~\bibnamefont
  {Braess}}, \bibinfo {author} {\bibfnamefont {A.}~\bibnamefont {Nagurney}},\
  and\ \bibinfo {author} {\bibfnamefont {T.}~\bibnamefont {Wakolbinger}},\
  }\bibfield  {title} {\bibinfo {title} {On a paradox of traffic planning},\
  }\href@noop {} {\bibfield  {journal} {\bibinfo  {journal} {Transportation
  science}\ }\textbf {\bibinfo {volume} {39}},\ \bibinfo {pages} {446}
  (\bibinfo {year} {2005})}\BibitemShut {NoStop}%
\bibitem [{\citenamefont {Sch{\"a}fer}\ \emph {et~al.}(2022)\citenamefont
  {Sch{\"a}fer}, \citenamefont {Pesch}, \citenamefont {Manik}, \citenamefont
  {Gollenstede}, \citenamefont {Lin}, \citenamefont {Beck}, \citenamefont
  {Witthaut},\ and\ \citenamefont {Timme}}]{schafer2022understanding}%
  \BibitemOpen
  \bibfield  {author} {\bibinfo {author} {\bibfnamefont {B.}~\bibnamefont
  {Sch{\"a}fer}}, \bibinfo {author} {\bibfnamefont {T.}~\bibnamefont {Pesch}},
  \bibinfo {author} {\bibfnamefont {D.}~\bibnamefont {Manik}}, \bibinfo
  {author} {\bibfnamefont {J.}~\bibnamefont {Gollenstede}}, \bibinfo {author}
  {\bibfnamefont {G.}~\bibnamefont {Lin}}, \bibinfo {author} {\bibfnamefont
  {H.-P.}\ \bibnamefont {Beck}}, \bibinfo {author} {\bibfnamefont
  {D.}~\bibnamefont {Witthaut}},\ and\ \bibinfo {author} {\bibfnamefont
  {M.}~\bibnamefont {Timme}},\ }\bibfield  {title} {\bibinfo {title}
  {Understanding braess’ paradox in power grids},\ }\href@noop {} {\bibfield
  {journal} {\bibinfo  {journal} {Nature Communications}\ }\textbf {\bibinfo
  {volume} {13}},\ \bibinfo {pages} {5396} (\bibinfo {year}
  {2022})}\BibitemShut {NoStop}%
\bibitem [{\citenamefont {Chandra}\ \emph {et~al.}(1989)\citenamefont
  {Chandra}, \citenamefont {Raghavan}, \citenamefont {Ruzzo},\ and\
  \citenamefont {Smolensky}}]{chandra1989electrical}%
  \BibitemOpen
  \bibfield  {author} {\bibinfo {author} {\bibfnamefont {A.~K.}\ \bibnamefont
  {Chandra}}, \bibinfo {author} {\bibfnamefont {P.}~\bibnamefont {Raghavan}},
  \bibinfo {author} {\bibfnamefont {W.~L.}\ \bibnamefont {Ruzzo}},\ and\
  \bibinfo {author} {\bibfnamefont {R.}~\bibnamefont {Smolensky}},\ }\bibfield
  {title} {\bibinfo {title} {The electrical resistance of a graph captures its
  commute and cover times},\ }in\ \href@noop {} {\emph {\bibinfo {booktitle}
  {Proceedings of the twenty-first annual ACM symposium on Theory of
  computing}}}\ (\bibinfo {year} {1989})\ pp.\ \bibinfo {pages}
  {574--586}\BibitemShut {NoStop}%
\bibitem [{\citenamefont {Wang}\ \emph
  {et~al.}(2017{\natexlab{b}})\citenamefont {Wang}, \citenamefont {Dubbeldam},\
  and\ \citenamefont {Van~Mieghem}}]{wang2017kemeny}%
  \BibitemOpen
  \bibfield  {author} {\bibinfo {author} {\bibfnamefont {X.}~\bibnamefont
  {Wang}}, \bibinfo {author} {\bibfnamefont {J.~L.}\ \bibnamefont
  {Dubbeldam}},\ and\ \bibinfo {author} {\bibfnamefont {P.}~\bibnamefont
  {Van~Mieghem}},\ }\bibfield  {title} {\bibinfo {title} {Kemeny's constant and
  the effective graph resistance},\ }\href@noop {} {\bibfield  {journal}
  {\bibinfo  {journal} {Linear Algebra and its Applications}\ }\textbf
  {\bibinfo {volume} {535}},\ \bibinfo {pages} {231} (\bibinfo {year}
  {2017}{\natexlab{b}})}\BibitemShut {NoStop}%
\bibitem [{\citenamefont {Kemeny}(1981)}]{kemeny1981generalization}%
  \BibitemOpen
  \bibfield  {author} {\bibinfo {author} {\bibfnamefont {J.~G.}\ \bibnamefont
  {Kemeny}},\ }\bibfield  {title} {\bibinfo {title} {Generalization of a
  fundamental matrix},\ }\href@noop {} {\bibfield  {journal} {\bibinfo
  {journal} {Linear Algebra and its Applications}\ }\textbf {\bibinfo {volume}
  {38}},\ \bibinfo {pages} {193} (\bibinfo {year} {1981})}\BibitemShut
  {NoStop}%
\bibitem [{\citenamefont {Levene}\ and\ \citenamefont
  {Loizou}(2002)}]{levene2002kemeny}%
  \BibitemOpen
  \bibfield  {author} {\bibinfo {author} {\bibfnamefont {M.}~\bibnamefont
  {Levene}}\ and\ \bibinfo {author} {\bibfnamefont {G.}~\bibnamefont
  {Loizou}},\ }\bibfield  {title} {\bibinfo {title} {Kemeny's constant and the
  random surfer},\ }\href@noop {} {\bibfield  {journal} {\bibinfo  {journal}
  {The American mathematical monthly}\ }\textbf {\bibinfo {volume} {109}},\
  \bibinfo {pages} {741} (\bibinfo {year} {2002})}\BibitemShut {NoStop}%
\bibitem [{\citenamefont {Hunter}(2014)}]{hunter2014role}%
  \BibitemOpen
  \bibfield  {author} {\bibinfo {author} {\bibfnamefont {J.~J.}\ \bibnamefont
  {Hunter}},\ }\bibfield  {title} {\bibinfo {title} {The role of kemeny's
  constant in properties of markov chains},\ }\href@noop {} {\bibfield
  {journal} {\bibinfo  {journal} {Communications in Statistics-Theory and
  Methods}\ }\textbf {\bibinfo {volume} {43}},\ \bibinfo {pages} {1309}
  (\bibinfo {year} {2014})}\BibitemShut {NoStop}%
\bibitem [{\citenamefont {Abdelnour}\ \emph {et~al.}(2014)\citenamefont
  {Abdelnour}, \citenamefont {Voss},\ and\ \citenamefont
  {Raj}}]{abdelnour2014network}%
  \BibitemOpen
  \bibfield  {author} {\bibinfo {author} {\bibfnamefont {F.}~\bibnamefont
  {Abdelnour}}, \bibinfo {author} {\bibfnamefont {H.~U.}\ \bibnamefont
  {Voss}},\ and\ \bibinfo {author} {\bibfnamefont {A.}~\bibnamefont {Raj}},\
  }\bibfield  {title} {\bibinfo {title} {Network diffusion accurately models
  the relationship between structural and functional brain connectivity
  networks},\ }\href@noop {} {\bibfield  {journal} {\bibinfo  {journal}
  {Neuroimage}\ }\textbf {\bibinfo {volume} {90}},\ \bibinfo {pages} {335}
  (\bibinfo {year} {2014})}\BibitemShut {NoStop}%
\bibitem [{\citenamefont {Mercier}\ \emph {et~al.}(2022)\citenamefont
  {Mercier}, \citenamefont {Scarpino},\ and\ \citenamefont
  {Moore}}]{mercier2022effective}%
  \BibitemOpen
  \bibfield  {author} {\bibinfo {author} {\bibfnamefont {A.}~\bibnamefont
  {Mercier}}, \bibinfo {author} {\bibfnamefont {S.}~\bibnamefont {Scarpino}},\
  and\ \bibinfo {author} {\bibfnamefont {C.}~\bibnamefont {Moore}},\ }\bibfield
   {title} {\bibinfo {title} {Effective resistance against pandemics: Mobility
  network sparsification for high-fidelity epidemic simulations},\ }\href@noop
  {} {\bibfield  {journal} {\bibinfo  {journal} {PLOS Computational Biology}\
  }\textbf {\bibinfo {volume} {18}},\ \bibinfo {pages} {e1010650} (\bibinfo
  {year} {2022})}\BibitemShut {NoStop}%
\bibitem [{\citenamefont {Horsevad}\ \emph {et~al.}(2022)\citenamefont
  {Horsevad}, \citenamefont {Mateo}, \citenamefont {Kooij}, \citenamefont
  {Barrat},\ and\ \citenamefont {Bouffanais}}]{horsevadtransition}%
  \BibitemOpen
  \bibfield  {author} {\bibinfo {author} {\bibfnamefont {N.}~\bibnamefont
  {Horsevad}}, \bibinfo {author} {\bibfnamefont {D.}~\bibnamefont {Mateo}},
  \bibinfo {author} {\bibfnamefont {R.~E.}\ \bibnamefont {Kooij}}, \bibinfo
  {author} {\bibfnamefont {A.}~\bibnamefont {Barrat}},\ and\ \bibinfo {author}
  {\bibfnamefont {R.}~\bibnamefont {Bouffanais}},\ }\bibfield  {title}
  {\bibinfo {title} {Transition from simple to complex contagion in collective
  decision-making},\ }\href@noop {} {\bibfield  {journal} {\bibinfo  {journal}
  {Nature communications}\ }\textbf {\bibinfo {volume} {13}},\ \bibinfo {pages}
  {1442} (\bibinfo {year} {2022})}\BibitemShut {NoStop}%
\bibitem [{\citenamefont {Zenil}\ \emph {et~al.}(2019)\citenamefont {Zenil},
  \citenamefont {Kiani}, \citenamefont {Zea},\ and\ \citenamefont
  {Tegn{\'e}r}}]{zenil2019causal}%
  \BibitemOpen
  \bibfield  {author} {\bibinfo {author} {\bibfnamefont {H.}~\bibnamefont
  {Zenil}}, \bibinfo {author} {\bibfnamefont {N.~A.}\ \bibnamefont {Kiani}},
  \bibinfo {author} {\bibfnamefont {A.~A.}\ \bibnamefont {Zea}},\ and\ \bibinfo
  {author} {\bibfnamefont {J.}~\bibnamefont {Tegn{\'e}r}},\ }\bibfield  {title}
  {\bibinfo {title} {Causal deconvolution by algorithmic generative models},\
  }\href@noop {} {\bibfield  {journal} {\bibinfo  {journal} {Nature Machine
  Intelligence}\ }\textbf {\bibinfo {volume} {1}},\ \bibinfo {pages} {58}
  (\bibinfo {year} {2019})}\BibitemShut {NoStop}%
\bibitem [{\citenamefont {Balwani}\ \emph {et~al.}(2025)\citenamefont
  {Balwani}, \citenamefont {Wang}, \citenamefont {Najafi},\ and\ \citenamefont
  {Choi}}]{balwani2025constructing}%
  \BibitemOpen
  \bibfield  {author} {\bibinfo {author} {\bibfnamefont {A.}~\bibnamefont
  {Balwani}}, \bibinfo {author} {\bibfnamefont {A.~Q.}\ \bibnamefont {Wang}},
  \bibinfo {author} {\bibfnamefont {F.}~\bibnamefont {Najafi}},\ and\ \bibinfo
  {author} {\bibfnamefont {H.}~\bibnamefont {Choi}},\ }\bibfield  {title}
  {\bibinfo {title} {Constructing biologically constrained rnns via dale’s
  backpropagation and topologically informed pruning},\ }\href@noop {}
  {\bibfield  {journal} {\bibinfo  {journal} {Science Advances}\ }\textbf
  {\bibinfo {volume} {11}},\ \bibinfo {pages} {eadw4970} (\bibinfo {year}
  {2025})}\BibitemShut {NoStop}%
\bibitem [{\citenamefont {Van~Mieghem}(2011)}]{VanMieghem2011}%
  \BibitemOpen
  \bibfield  {author} {\bibinfo {author} {\bibfnamefont {P.}~\bibnamefont
  {Van~Mieghem}},\ }\href@noop {} {\emph {\bibinfo {title} {Graph spectra for
  complex networks}}}\ (\bibinfo  {publisher} {Cambridge University Press},\
  \bibinfo {year} {2011})\BibitemShut {NoStop}%
\end{thebibliography}

\begin{thebibliography}{8}%
\makeatletter
\providecommand \@ifxundefined [1]{%
 \@ifx{#1\undefined}
}%
\providecommand \@ifnum [1]{%
 \ifnum #1\expandafter \@firstoftwo
 \else \expandafter \@secondoftwo
 \fi
}%
\providecommand \@ifx [1]{%
 \ifx #1\expandafter \@firstoftwo
 \else \expandafter \@secondoftwo
 \fi
}%
\providecommand \natexlab [1]{#1}%
\providecommand \enquote  [1]{``#1''}%
\providecommand \bibnamefont  [1]{#1}%
\providecommand \bibfnamefont [1]{#1}%
\providecommand \citenamefont [1]{#1}%
\providecommand \href@noop [0]{\@secondoftwo}%
\providecommand \href [0]{\begingroup \@sanitize@url \@href}%
\providecommand \@href[1]{\@@startlink{#1}\@@href}%
\providecommand \@@href[1]{\endgroup#1\@@endlink}%
\providecommand \@sanitize@url [0]{\catcode `\\12\catcode `\$12\catcode
  `\&12\catcode `\#12\catcode `\^12\catcode `\_12\catcode `\%12\relax}%
\providecommand \@@startlink[1]{}%
\providecommand \@@endlink[0]{}%
\providecommand \url  [0]{\begingroup\@sanitize@url \@url }%
\providecommand \@url [1]{\endgroup\@href {#1}{\urlprefix }}%
\providecommand \urlprefix  [0]{URL }%
\providecommand \Eprint [0]{\href }%
\providecommand \doibase [0]{https://doi.org/}%
\providecommand \selectlanguage [0]{\@gobble}%
\providecommand \bibinfo  [0]{\@secondoftwo}%
\providecommand \bibfield  [0]{\@secondoftwo}%
\providecommand \translation [1]{[#1]}%
\providecommand \BibitemOpen [0]{}%
\providecommand \bibitemStop [0]{}%
\providecommand \bibitemNoStop [0]{.\EOS\space}%
\providecommand \EOS [0]{\spacefactor3000\relax}%
\providecommand \BibitemShut  [1]{\csname bibitem#1\endcsname}%
\let\auto@bib@innerbib\@empty
\bibitem [{\citenamefont {Lusseau}\ \emph {et~al.}(2003)\citenamefont
  {Lusseau}, \citenamefont {Schneider}, \citenamefont {Boisseau}, \citenamefont
  {Haase}, \citenamefont {Slooten},\ and\ \citenamefont
  {Dawson}}]{lusseau2003bottlenose}%
  \BibitemOpen
  \bibfield  {author} {\bibinfo {author} {\bibfnamefont {D.}~\bibnamefont
  {Lusseau}}, \bibinfo {author} {\bibfnamefont {K.}~\bibnamefont {Schneider}},
  \bibinfo {author} {\bibfnamefont {O.~J.}\ \bibnamefont {Boisseau}}, \bibinfo
  {author} {\bibfnamefont {P.}~\bibnamefont {Haase}}, \bibinfo {author}
  {\bibfnamefont {E.}~\bibnamefont {Slooten}},\ and\ \bibinfo {author}
  {\bibfnamefont {S.~M.}\ \bibnamefont {Dawson}},\ }\bibfield  {title}
  {\bibinfo {title} {The bottlenose dolphin community of doubtful sound
  features a large proportion of long-lasting associations: can geographic
  isolation explain this unique trait?},\ }\href@noop {} {\bibfield  {journal}
  {\bibinfo  {journal} {Behavioral ecology and sociobiology}\ }\textbf
  {\bibinfo {volume} {54}},\ \bibinfo {pages} {396} (\bibinfo {year}
  {2003})}\BibitemShut {NoStop}%
\bibitem [{\citenamefont {Leskovec}\ \emph {et~al.}(2007)\citenamefont
  {Leskovec}, \citenamefont {Kleinberg},\ and\ \citenamefont
  {Faloutsos}}]{leskovec2007graph}%
  \BibitemOpen
  \bibfield  {author} {\bibinfo {author} {\bibfnamefont {J.}~\bibnamefont
  {Leskovec}}, \bibinfo {author} {\bibfnamefont {J.}~\bibnamefont
  {Kleinberg}},\ and\ \bibinfo {author} {\bibfnamefont {C.}~\bibnamefont
  {Faloutsos}},\ }\bibfield  {title} {\bibinfo {title} {Graph evolution:
  Densification and shrinking diameters},\ }\href@noop {} {\bibfield  {journal}
  {\bibinfo  {journal} {ACM transactions on Knowledge Discovery from Data
  (TKDD)}\ }\textbf {\bibinfo {volume} {1}},\ \bibinfo {pages} {2} (\bibinfo
  {year} {2007})}\BibitemShut {NoStop}%
\bibitem [{\citenamefont {Chami}\ \emph {et~al.}(2017)\citenamefont {Chami},
  \citenamefont {Ahnert}, \citenamefont {Kabatereine},\ and\ \citenamefont
  {Tukahebwa}}]{chami2017social}%
  \BibitemOpen
  \bibfield  {author} {\bibinfo {author} {\bibfnamefont {G.~F.}\ \bibnamefont
  {Chami}}, \bibinfo {author} {\bibfnamefont {S.~E.}\ \bibnamefont {Ahnert}},
  \bibinfo {author} {\bibfnamefont {N.~B.}\ \bibnamefont {Kabatereine}},\ and\
  \bibinfo {author} {\bibfnamefont {E.~M.}\ \bibnamefont {Tukahebwa}},\
  }\bibfield  {title} {\bibinfo {title} {Social network fragmentation and
  community health},\ }\href@noop {} {\bibfield  {journal} {\bibinfo  {journal}
  {Proceedings of the National Academy of Sciences}\ }\textbf {\bibinfo
  {volume} {114}},\ \bibinfo {pages} {E7425} (\bibinfo {year}
  {2017})}\BibitemShut {NoStop}%
\bibitem [{\citenamefont {Guimera}\ \emph {et~al.}(2003)\citenamefont
  {Guimera}, \citenamefont {Danon}, \citenamefont {Diaz-Guilera}, \citenamefont
  {Giralt},\ and\ \citenamefont {Arenas}}]{guimera2003self}%
  \BibitemOpen
  \bibfield  {author} {\bibinfo {author} {\bibfnamefont {R.}~\bibnamefont
  {Guimera}}, \bibinfo {author} {\bibfnamefont {L.}~\bibnamefont {Danon}},
  \bibinfo {author} {\bibfnamefont {A.}~\bibnamefont {Diaz-Guilera}}, \bibinfo
  {author} {\bibfnamefont {F.}~\bibnamefont {Giralt}},\ and\ \bibinfo {author}
  {\bibfnamefont {A.}~\bibnamefont {Arenas}},\ }\bibfield  {title} {\bibinfo
  {title} {Self-similar community structure in a network of human
  interactions},\ }\href@noop {} {\bibfield  {journal} {\bibinfo  {journal}
  {Physical review E}\ }\textbf {\bibinfo {volume} {68}},\ \bibinfo {pages}
  {065103} (\bibinfo {year} {2003})}\BibitemShut {NoStop}%
\bibitem [{\citenamefont {Crucitti}\ \emph {et~al.}(2006)\citenamefont
  {Crucitti}, \citenamefont {Latora},\ and\ \citenamefont
  {Porta}}]{crucitti2006centrality}%
  \BibitemOpen
  \bibfield  {author} {\bibinfo {author} {\bibfnamefont {P.}~\bibnamefont
  {Crucitti}}, \bibinfo {author} {\bibfnamefont {V.}~\bibnamefont {Latora}},\
  and\ \bibinfo {author} {\bibfnamefont {S.}~\bibnamefont {Porta}},\ }\bibfield
   {title} {\bibinfo {title} {Centrality measures in spatial networks of urban
  streets},\ }\href@noop {} {\bibfield  {journal} {\bibinfo  {journal}
  {Physical Review E—Statistical, Nonlinear, and Soft Matter Physics}\
  }\textbf {\bibinfo {volume} {73}},\ \bibinfo {pages} {036125} (\bibinfo
  {year} {2006})}\BibitemShut {NoStop}%
\bibitem [{\citenamefont {Cardillo}\ \emph {et~al.}(2013)\citenamefont
  {Cardillo}, \citenamefont {G{\'o}mez-Gardenes}, \citenamefont {Zanin},
  \citenamefont {Romance}, \citenamefont {Papo}, \citenamefont {Pozo},\ and\
  \citenamefont {Boccaletti}}]{cardillo2013emergence}%
  \BibitemOpen
  \bibfield  {author} {\bibinfo {author} {\bibfnamefont {A.}~\bibnamefont
  {Cardillo}}, \bibinfo {author} {\bibfnamefont {J.}~\bibnamefont
  {G{\'o}mez-Gardenes}}, \bibinfo {author} {\bibfnamefont {M.}~\bibnamefont
  {Zanin}}, \bibinfo {author} {\bibfnamefont {M.}~\bibnamefont {Romance}},
  \bibinfo {author} {\bibfnamefont {D.}~\bibnamefont {Papo}}, \bibinfo {author}
  {\bibfnamefont {F.~d.}\ \bibnamefont {Pozo}},\ and\ \bibinfo {author}
  {\bibfnamefont {S.}~\bibnamefont {Boccaletti}},\ }\bibfield  {title}
  {\bibinfo {title} {Emergence of network features from multiplexity},\
  }\href@noop {} {\bibfield  {journal} {\bibinfo  {journal} {Scientific
  reports}\ }\textbf {\bibinfo {volume} {3}},\ \bibinfo {pages} {1344}
  (\bibinfo {year} {2013})}\BibitemShut {NoStop}%
\bibitem [{\citenamefont {Leskovec}\ and\ \citenamefont
  {Mcauley}(2012)}]{leskovec2012learning}%
  \BibitemOpen
  \bibfield  {author} {\bibinfo {author} {\bibfnamefont {J.}~\bibnamefont
  {Leskovec}}\ and\ \bibinfo {author} {\bibfnamefont {J.}~\bibnamefont
  {Mcauley}},\ }\bibfield  {title} {\bibinfo {title} {Learning to discover
  social circles in ego networks},\ }\href@noop {} {\bibfield  {journal}
  {\bibinfo  {journal} {Advances in neural information processing systems}\
  }\textbf {\bibinfo {volume} {25}} (\bibinfo {year} {2012})}\BibitemShut
  {NoStop}%
\bibitem [{\citenamefont {Sapiezynski}\ \emph {et~al.}(2019)\citenamefont
  {Sapiezynski}, \citenamefont {Stopczynski}, \citenamefont {Lassen},\ and\
  \citenamefont {Lehmann}}]{sapiezynski2019interaction}%
  \BibitemOpen
  \bibfield  {author} {\bibinfo {author} {\bibfnamefont {P.}~\bibnamefont
  {Sapiezynski}}, \bibinfo {author} {\bibfnamefont {A.}~\bibnamefont
  {Stopczynski}}, \bibinfo {author} {\bibfnamefont {D.~D.}\ \bibnamefont
  {Lassen}},\ and\ \bibinfo {author} {\bibfnamefont {S.}~\bibnamefont
  {Lehmann}},\ }\bibfield  {title} {\bibinfo {title} {Interaction data from the
  copenhagen networks study},\ }\href@noop {} {\bibfield  {journal} {\bibinfo
  {journal} {Scientific Data}\ }\textbf {\bibinfo {volume} {6}},\ \bibinfo
  {pages} {315} (\bibinfo {year} {2019})}\BibitemShut {NoStop}%
\end{thebibliography}
